\documentclass[10pt,twocolumn,reprint,amsmath,amssymb,aps,pra,showpacs,longbibliography,superscriptaddress]{revtex4-1}
\usepackage[latin9]{inputenc}
\usepackage{amsmath}
\usepackage{amssymb}
\usepackage{graphicx}

\makeatletter

\usepackage{amsfonts}
\usepackage{graphics}

\makeatother

\begin{document}
\title{Functional Determinant Approach Investigations of Heavy Impurity Physics}
\author{Jia Wang}
\affiliation{Centre for Quantum Technology Theory, Swinburne University of Technology,
Melbourne 3122, Australia}
\date{\today}
\begin{abstract}
In this brief review, we report some new development in the functional
determinant approach (FDA), an exact numerical method, in the studies
of a heavy quantum impurity immersed in Fermi gases and manipulated
with radio-frequency pulses. FDA has been successfully applied to
investigate the universal dynamical responses of a heavy impurity
in an ultracold ideal Fermi gas in both the time and frequency domain,
which allows the exploration of the renowned Anderson's orthogonality
catastrophe (OC). In such a system, OC is induced by the multiple
particle-hole excitations of the Fermi sea, which is beyond a simple
perturbation picture and manifests itself as the absence of quasiparticles
named polarons. More recently, two new directions for studying heavy
impurity with FDA have been developed. One is to extend FDA to a strongly
correlated background superfluid background, a Bardeen--Cooper--Schrieffer
(BCS) superfluid. In this system, Anderson's orthogonality catastrophe
is prohibited due to the suppression of multiple particle-hole excitations
by the superfluid gap, which leads to the existence of genuine polaron.
The other direction is to generalize the FDA to the case of multiple
RF pulses scheme, which extends the well-established 1D Ramsey spectroscopy
in ultracold atoms into multidimensional, in the same spirit as the
well-known multidimensional nuclear magnetic resonance and optical
multidimensional coherent spectroscopy. Multidimensional Ramsey spectroscopy
allows us to investigate correlations between spectral peaks of an
impurity-medium system that is not accessible in the conventional
one-dimensional spectrum.
\end{abstract}
\maketitle

\section{Introduction}

An important approach to investigating polaron physics is to study
the heavy impurity limit. Infinitely heavy impurity interacting with
a Fermi sea represents one of the rare examples of exactly solvable
many-body problems in the nonperturbative regime, which can serve
as a benchmark for various approximations. Historically, this problem
originated from the studies of the x-ray spectra in metals, where
Mahan predicts the so-called Fermi edge singularities (FES), absorption
edges in the spectra characterized by a power law divergence near
the threshold \cite{Mahan2000Book}. The optical transition is determined
by a highly spatial localized core-level hole that can be regarded
as an impurity with infinite mass immersed in a Fermi sea of conduction
electrons. The corresponding model Hamiltonian can be solved exactly
and is often called MND Hamiltonian in the condensed matter community
after the work of Mahan \cite{Mahan1967PR1,Mahan1967PR2} and Nozi{\'e}res-De
Dominicis \cite{Nozieres1969PR}.

FES is the first and one of the most important examples of nonequilibrium
many-body physics. The underlying physics can be interpreted by the
concept of Anderson's orthogonality catastrophe (OC) \cite{Anderson1967PRL},
i.e., the many-particle states with and without impurity become orthogonal.
FES has also been observed in current-voltage characteristics of resonant
tunneling experiments dominated by localized states \cite{Matveev1992PRB,Pate1994PRL}
and has been proposed to be investigated in various systems, including
quantum wires \cite{Nagaosa1992PRL,Prokof'ev1994PRB,Komnik1997PRB},
and quantum dots \cite{bascones2000PRB}. In particular, a convenient
and numerically exact method, namely the functional determinant approach
(FDA) \cite{Leonid1996JMathPhys,Klich2003Book,Schonhammer2007PRB,Ivanov2013JMathPhys},
has been developed to study FES in out-of-equilibrium Fermi gases
\cite{Muzykantskii2003PRL,Muzykantskii2005PRB,Levitov2005PRL} and
open quantum dots \cite{Levitov2004PRL}. Using FDA to investigate
MND Hamiltonians has also been applied to study exciton-polarons in
monolayer transition metal dichalcogenides (TMD), where the exciton
serves as the impurity, and the itinerant excess electrons play the
role of the background Fermi sea \cite{Reichman2019PRB,Reichman2022arXiv}.
However, the prediction can only be considered qualitative here, as
the exciton mass is only about twice the electron mass in TMDs.

In recent years, ultracold quantum gases have emerged as an ideal
testbed for impurity physics thanks to their unprecedented controllability.
In the context of ultracold Fermi gases, the FES of an infinitely
heavy impurity in an ideal Fermi gas has been quantitatively re-examined
via the FDA \cite{Demler2012PRX,Schmidt2018Review} and can be verified
via Ramsey-interference-type experiments \cite{Goold2011PRA}. The
Ramsey signals in the time domain are universal, i.e., fully determined
by the impurity-medium scattering length and the Fermi wave vector
of the medium Fermi gases, not only in the long-time limit (as their
counterpart in solid-state systems) but also for all times. Corresponding
spectra in the frequency domain obtained by Fourier transformation
show FES and provide an insightful understanding of polaron physics.
The exact results of the FDA can serve as benchmark calculations for
various approximation calculations of Fermi polarons, such as Chevy's
ansatz or equivalently many-body T-matrix \cite{Chevy2006PRA,Combescot2007PRL,Punk2009PRA,Cui2010PRA,Mathy2011PRL,Schmidt2012PRA,Parish2013PRA,Levinsen2015PRL,HuHui2016PRA,HuHui2018PRA,Mulkerin2019AnnPhys,Parish2021PRA},
and other exact methods, such as quantum Monte Carlo methods \cite{Lobo2006PRL,Kroiss2015PRL,Goulko2016PRA,Pessoa2021PRA}.

However, polaron, strictly speaking, does not exist in the infinitely
heavy impurity limit, where the quasiparticle residue of polaron vanishes
due to the presence of OC \cite{Demler2012PRX,Schmidt2018Review}.
On the other hand, the generalization of the FDA to finite impurity
mass remains elusive. Nevertheless, FDA has been proven to be qualitatively
accurate in describing heavy polarons in ultracold Fermi gases at
a finite temperature, where thermal fluctuation is comparable with
recoil energy \cite{Demler2016Science,Meera2019PRL}. In addition,
one can choose an impurity with very different polarizability from
the background fermions. As a result, the impurity can be confined
by a deep optical lattice or an optical tweezer without affecting
the itinerant background fermions. In this case, the infinitely heavy
mass limit becomes exact, and FDA calculations can serve as a critical
meeting point for theoretical and experimental efforts to understand
the complicated quantum dynamics of interacting many-particle systems.
Inspired by the pioneer works \cite{Demler2012PRX,Schmidt2018Review},
a heavy impurity in Fermi gases has also been proposed to investigate
spin transportation \cite{Demler2019PRB} and precise measurement
of the temperature of noninteracting Fermi gases \cite{Goold2020PRL}.
The exact finite-temperature free energy and Tan contact \cite{Braaten2010PRL},
as well as the exact dynamics of Tan contact of a heavy impurity in
ideal Fermi gases, can be derived as a generalization of FDA \cite{Meera2020PRA}.
Rabi oscillations of heavy impurities in an ideal Fermi gas can also
be investigated via FDA \cite{Meera2021PRA}. Extensions of the FDA
to the investigation of Rydberg impurities \cite{Pfau2013Nature,Jia2015PRL}
in Fermi \cite{Schmidt2020PRR} and Bose gases \cite{Schmidt2016PRL,Camargo2018PRL,Schmidt2018PRA}
have also been developed recently.

Here, we briefly review the formalism of the FDA and two recent developments.
Firstly, FDA has been generalized to the system of a heavy impurity
in a Bardeen--Cooper--Schrieffer (BCS) superfluid, where the strongly
correlated superfluid background is described by a BCS mean-field
wavefunction \cite{JiaWang2022PRLshort,JiaWang2022PRAlong}. In contrast
to the ideal Fermi gas case, the pairing gap in the BCS superfluid
prevents the OC and leads to genuine polaron signals in the spectrum
even at zero temperature. In addition, at finite temperature, additional
features related to the subgap Yu-Shiba-Rusinov (YSR) bound state
ware predicted in the spectra of a magnetic impurity. Another recent
development is to extend the FDA to multidimensional (MD) spectroscopy.
In contrast to conventional one-dimensional (1D) spectroscopy which
depends only on one variable, such as photon frequency, MD spectroscopy
unfolds spectral information into several dimensions, revealing correlations
between spectral peaks that the 1D spectrum cannot access.

\section{Formalism}

\subsection{System setup}

\begin{figure}
\includegraphics[width=0.98\columnwidth]{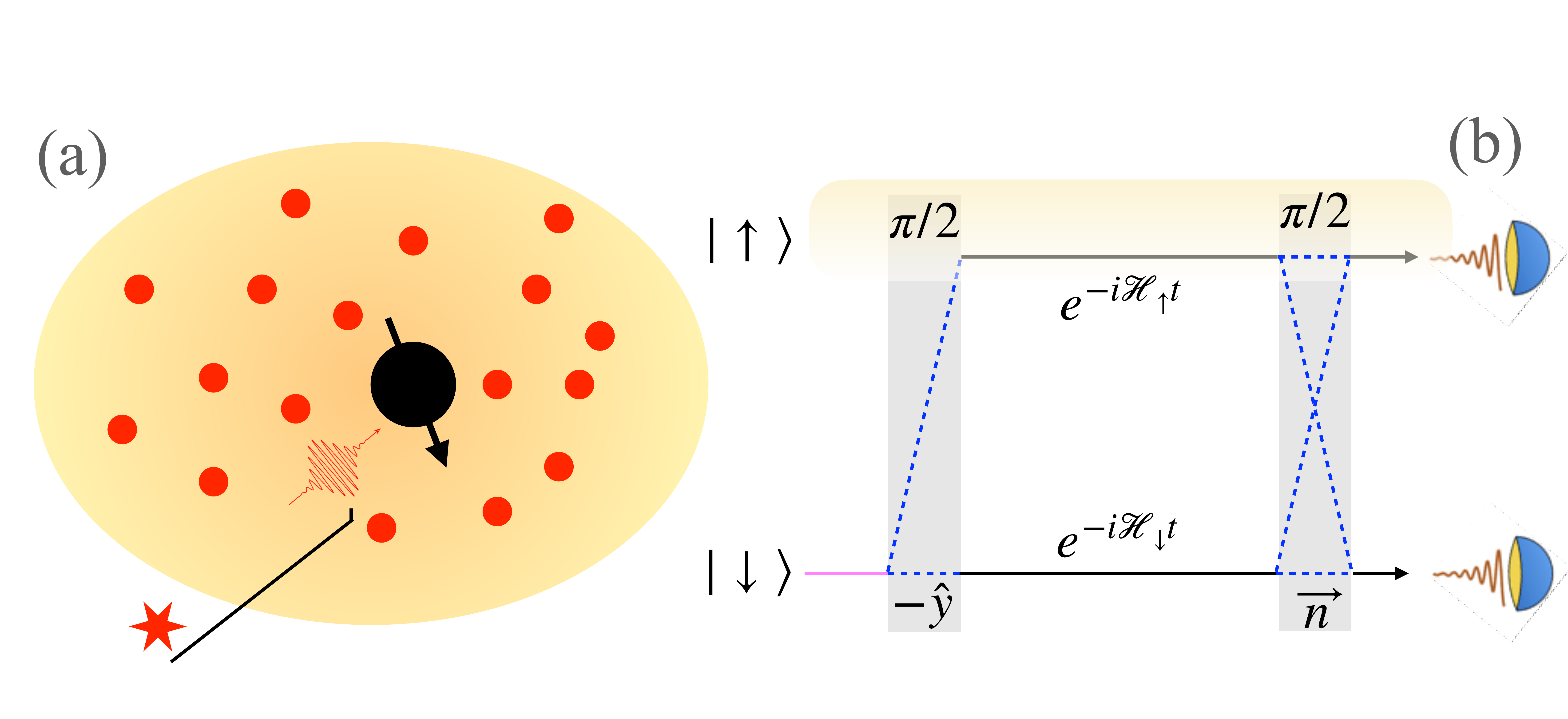}

\caption{(a) A sketch of the system setup for 1D Ramsey spectroscopy (injection
scheme). (b) An interferometry interpretation of 1D Ramsey spectroscopy.
\label{fig:Sketch1D}}

\end{figure}

The basic setup of our system is shown in Fig. \ref{fig:Sketch1D}
(a). We place a localized fermionic or bosonic impurity (the big black
ball with a black arrow) with two internal pseudospins states $|\uparrow\rangle$
and $|\downarrow\rangle$ , which we will call spins for short from
now on, in the background of ultracold Fermi gas (the red dots). In
real experiments, there is usually more than one impurity, but the
impurity density is prepared to be very low so that the interaction
between impurities can be regarded as negligible. As mentioned before,
the localization of impurity can be either achieved by confinement
of a deep optical trap or treated as an approximation to an impurity
atom with heavy mass. Unless specified otherwise, we are interested
in the case where the interaction between the background Fermi gas
and $|\downarrow\rangle$ is negligible, while the interaction with
$|\uparrow\rangle$ is arbitrarily tunable by, e.g., Feshbach resonances.
(It is straightforward to generalize to the case where both $|\uparrow\rangle$
and $|\downarrow\rangle$ interact with the background.) The spins
states can be manipulated by radio-frequency (RF) pulses, which assume
to be able to rotate the spins infinitely fast. In reality, the RF
pulse length is usually comparable with the characteristic time scale
$\tau_{F}=E_{F}^{-1}$ of the background Fermi gases, where $E_{F}$
is the Fermi energy, and we use unit $\hbar=1$ throughout this work.
For example, in Ref. \cite{Demler2016Science}, the typical pulse
length is about $10$ $\mu$s, approximately $3.4$ $\tau_{F}$ in
their system. However, the optical control of Feshbach resonances
in their experiment can be achieved very rapidly in less than $200$
ns, which is about 0.08 $\tau_{F}$. As a result, one can switch off
the interaction (for both spin states) in no time and rotate the spin
without perturbing the background Fermi gas, which can be treated
as an infinitely fast rotation theoretically. The interaction is switched
back on after the rotation. In principle, one can rotate the spin
in the Block sphere along an arbitrary axis, characterized by a unit
vector $\vec{n}=(n_{x},n_{y},n_{z})$, for an arbitrary angle $\theta$.
The rotation can be described by a unitary matrix in the spin basis
as
\begin{equation}
R_{\vec{n}}(\theta)=\exp(-i\frac{\theta}{2}\vec{n}\cdot\vec{\sigma}),
\end{equation}
where $\vec{\sigma}=(\sigma_{x},\sigma_{y},\sigma_{z})$ and $\sigma_{x}$,
$\sigma_{y}$, and $\sigma_{z}$ are Pauli matrices in the spin basis.
A $\pi/2$-pulse along the $-\hat{y}$-axis gives $R_{-\hat{y}}(\pi/2)|\downarrow\rangle=(|\uparrow\rangle+|\downarrow\rangle)/\sqrt{2}$.

The basic 1D Ramsey interferometric can be intuitively understood
by the sketch in Fig. \ref{fig:Sketch1D} (b). The effectively infinitely
fast rotation allows one to prepare the system in a superposition
state $|\Psi(0)\rangle=|\psi_{{\rm FS}}\rangle\otimes(|\uparrow\rangle+|\downarrow\rangle)/\sqrt{2}$,
where $|\psi_{{\rm FS}}\rangle$ describes the zero-temperature ground
state of the Fermi gas. For a single component Fermi gas, $|\psi_{{\rm FS}}\rangle$
corresponds to all fermions occupying the lowest eigenenergy states,
i.e., a Fermi sea. Here, we first briefly describe the general idea
using pure and zero-temperature states. The detailed formalization
and the straightforward generalization to finite-temperature density
matrix description will be given later.

Since the two spin states interact differently with the Fermi sea,
the associated time evolution operator after time $t$ are different:

\begin{equation}
|\Psi(t)\rangle=\frac{1}{\sqrt{2}}(|\uparrow\rangle\otimes e^{-i\mathcal{H}_{\uparrow}t}|\psi_{{\rm FS}}\rangle+|\downarrow\rangle\otimes e^{-i\mathcal{H}_{\downarrow}t}|\psi_{{\rm FS}}\rangle),
\end{equation}
where $\mathcal{H}_{\uparrow}$ and $\mathcal{H}_{\downarrow}$ are
the Hamiltonian for a Fermi sea with an interacting and noninteracting
impurity, respectively. The so-called many-body overlap function
\begin{equation}
S(t)\equiv\langle\psi_{{\rm FS}}|e^{i\mathcal{H}_{\downarrow}t}e^{-i\mathcal{H}_{\uparrow}t}|\psi_{{\rm FS}}\rangle\label{eq:1DRamsey0T}
\end{equation}
can be measured via the interference
\begin{equation}
{\rm Re}S(t)=\langle\sigma_{x}\rangle,\ {\rm Im}S(t)=-\langle\sigma_{y}\rangle,
\end{equation}
or equivalently $S(t)=\langle\sigma_{-}\rangle$ with $\sigma_{-}=\sigma_{x}-i\sigma_{y}$.
Notice that for non-interacting $|\downarrow\rangle$, $\mathcal{H}_{\downarrow}|\psi_{{\rm FS}}\rangle=E_{{\rm FS}}|\psi_{{\rm FS}}\rangle$
with $E_{{\rm FS}}$ being the Fermi sea energy, i.e., the summation
of eigenenergies of occupied states. Consequently, the overlap function
$S(t)=e^{iE_{{\rm FS}}t}\langle\psi_{{\rm FS}}|e^{-i\mathcal{H}_{\uparrow}t}|\psi_{{\rm FS}}\rangle$
takes the form of Loschmidt amplitude, the central object within the
theory of dynamical quantum phase transitions \cite{Heyl2018Rpp}.

A direct measurement of $\langle\sigma_{x}\rangle$ and $-\langle\sigma_{y}\rangle$
might not be as convenient as measuring $\langle\sigma_{z}\rangle=\left(N_{\uparrow}-N_{\downarrow}\right)/(N_{\uparrow}+N_{\downarrow})$,
where $N_{\uparrow}$ and $N_{\downarrow}$ are the population of
spin-up and spin-down impurities, respectively. (As mentioned above,
there are usually a finite number of independent impurities in a realistic
experiment.) Consequently, a standard protocol is to perform another
rotation after the evolution time $t$. From the relation$R_{-\hat{y}}(\pi/2)^{-1}\sigma_{z}R_{-\hat{y}}(\pi/2)=\sigma_{x}$
and $R_{-\hat{x}}(\pi/2)^{-1}\sigma_{z}R_{-\hat{x}}(\pi/2)=-\sigma_{y}$,
we can see that $\langle\sigma_{x}\rangle$ and $-\langle\sigma_{y}\rangle$
can be obtained by measuring $\sigma_{z}$ after rotation $R_{-\hat{y}}(\pi/2)$
and $R_{-\hat{x}}(\pi/2)$, respectively.

Since $|\uparrow\rangle$ and $|\downarrow\rangle$ correspond equivalently
to the existence and absence of impurity in the single impurity case,
we can therefore construct the creation operator $\hat{b}^{\dagger}$
and annihilation operator $\hat{b}$ for the impurity so that the
full Hamiltonian can be written as

\begin{equation}
\mathcal{\hat{H}}=\mathcal{\hat{H}}_{\uparrow}|\uparrow\rangle\langle\uparrow|+\mathcal{\hat{H}}_{\downarrow}|\downarrow\rangle\langle\downarrow|=\mathcal{\hat{H}}_{\uparrow}\hat{b}^{\dagger}\hat{b}+\mathcal{\hat{H}}_{\downarrow}(1-\hat{b}^{\dagger}\hat{b}).\label{eq:HamiltonianForm}
\end{equation}
The retarded Green's function for the impurity can thus be written
as
\begin{equation}
G_{I}(t)=-i\Theta(t)\langle\hat{b}(t)\hat{b}^{\dagger}\rangle,
\end{equation}
where $\hat{b}(t)=e^{i\mathcal{\hat{H}}t}\hat{b}e^{-i\mathcal{\hat{H}}t}$
in the Heisenberg picture. Tracing out the spin degree of freedom,
we have the relationship between the retarded Green's function and
the many-body overlap function as
\begin{equation}
G_{I}(t)=-iS(t),\ t>0.
\end{equation}
As a result, the Fourier transformation 
\begin{equation}
A(\omega)=\frac{1}{\pi}\int_{0}^{\infty}e^{i\omega t}S(t)dt=\frac{i}{\pi}\mathcal{G}_{I}(\omega),
\end{equation}
is related to the retarded Green's function in the frequency domain
$\mathcal{G}_{I}(\omega)=\int_{0}^{\infty}e^{i\omega t}G_{I}(t)dt$,
where the spectral function ${\rm Re}A(\omega)=-{\rm Im}\mathcal{G}_{I}(\omega)/\pi$
gives the absorption spectrum in the linear response regime. Throughout
this work, ${\rm Re}$ and ${\rm Im}$ denote the real and imaginary
parts of a complex number, respectively.

\subsection{Functional Determinant Approach \label{subsec:FDA}}

\begin{figure}
\includegraphics[width=0.98\columnwidth]{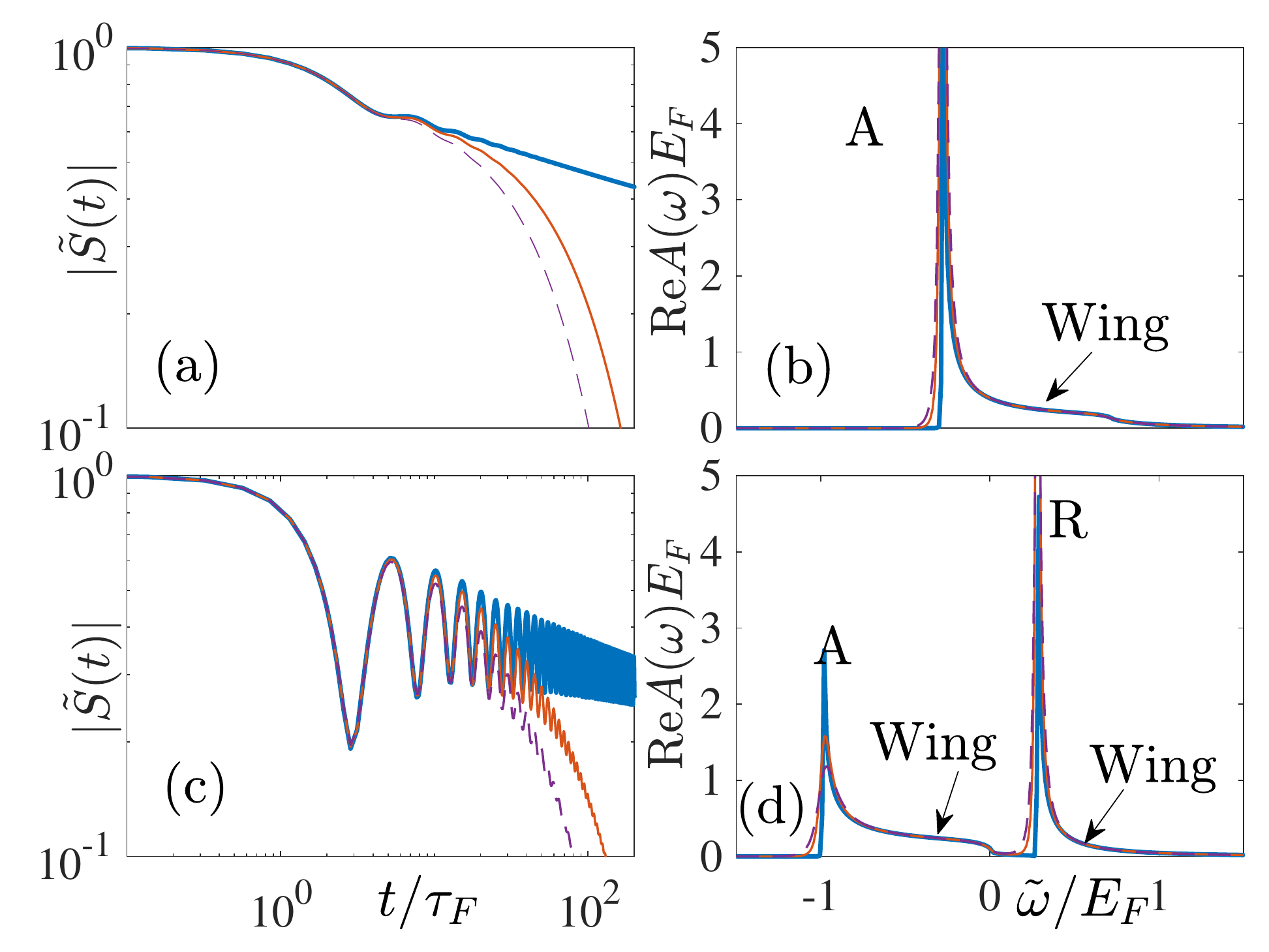}

\caption{1D Ramsey spectroscopy for (a) (b) attractive interaction $k_{F}a=-2$
and (c) (d) repulsive interaction $k_{F}a=2$. (a) and (c) show the
overlap functions $\tilde{S}(t)$. (b) and (d) show the spectral functions
${\rm Re}A(\omega)$. Thick blue curves correspond to $k_{B}T^{\circ}=0$,
thin red solid curves, and purple dashed curves correspond to $k_{B}T^{\circ}=0.03E_{F}$
and $k_{B}T^{\circ}=0.05E_{F}$, respectively. \label{fig:Ramsey1DFES}}
\end{figure}

In the previous section, we have given a general discussion of the
underlying idea of 1D Ramsey responses of a heavy impurity and its
relation to the absorption spectrum. Here, we show the detail of how
to exactly solve the time-dependent problem nonperturbatively using
the FDA. As a concrete example, we focus on the case where the background
is a dilute single-component Fermi gas, which is considered to be
noninteracting at ultracold temperature due to Pauli's exclusion principle.
As mentioned before, we assume only the background fermion only interacts
with $|\uparrow\rangle$, which is dominated by the $s$-wave interaction
that can be tuned via, e.g., Feshbach resonances. The corresponding
Hamiltonian can be expressed in the form of Eq. \ref{eq:HamiltonianForm},
where

\begin{equation}
\mathcal{\hat{H}}_{\uparrow}=\mathcal{\hat{H}}_{\downarrow}+\sum_{\mathbf{k},\mathbf{q}}\tilde{V}(\mathbf{k}-\mathbf{q})\hat{c}_{\mathbf{k}}^{\dagger}\hat{c}_{\mathbf{q}}+\omega_{s},\ \mathcal{\hat{H}}_{\downarrow}=\sum_{\mathbf{k}}\epsilon_{\mathbf{k}}\hat{c}_{\mathbf{k}}^{\dagger}\hat{c}_{\mathbf{k}}.\label{eq:Hamiltonian_idealfermion}
\end{equation}
Here, $\omega_{s}$ denotes the energy differences between the two
spin levels. $\hat{c}_{\mathbf{k}}^{\dagger}$ and $\hat{c}_{\mathbf{k}}$
are creation and annihilation operators of the background fermions
with momentum $\mathbf{k}$, respectively. $\epsilon_{\mathbf{k}}=k^{2}/2m$
is the single-particle kinetic energy of the background fermions with
mass $m$. $\tilde{V}(\mathbf{k})$ is the Fourier transform of $V(\mathbf{r})$,
the interaction potential between $|\uparrow\rangle$ and the background
fermions. The low-temperature physics is determined by the $s$-wave
energy-dependent scattering length $a(E_{F})=-\tan\eta(k_{F})/k_{F}$
at the Fermi energy $E_{F}=k_{F}^{2}/2m$, with $\eta(E_{F})$ being
an energy-dependent $s$-wave scattering phase-shift obtained from
a two-body scattering calculation with potential $V(\mathbf{r})$.
For the simplicity of notation, we denote $a\equiv a(E_{F})$ hereafter.

For the example given here, we are interested in the so-called injection
scheme where the spin is initially prepared in the noninteracting
state $|\downarrow\rangle$. The initial density matrix of the system
can therefore be written as $\rho_{i}=\rho_{{\rm FS}}\otimes|\downarrow\rangle\langle\downarrow|$,
where the thermal density matrix of the background fermion at a finite
temperature $T^{\circ}$ is given by 
\begin{equation}
\rho_{{\rm FS}}=\prod_{\mathbf{k}}\left[n_{\mathbf{k}}\hat{c}_{\mathbf{k}}^{\dagger}\hat{c}_{\mathbf{k}}+\left(1-n_{\mathbf{k}}\right)\hat{c}_{\mathbf{k}}\hat{c}_{\mathbf{k}}^{\dagger}\right],
\end{equation}
with the occupation of the momentum state
\begin{equation}
n_{\mathbf{k}}=\frac{1}{e^{\left(\epsilon_{\mathbf{k}}-\mu\right)/k_{B}T^{\circ}}+1}.
\end{equation}
Here, $k_{B}$ is the Boltzman constant, $\mu\simeq E_{F}$ is the
chemical potential determined by the number density of the background
Fermi gas. We also define a diagonal matrix $\hat{n}$ with the matrix
elements $n_{\mathbf{k}}$, which will become useful later.

For the simple 1D Ramsey spectrum, we apply a $\pi/2$ RF pulse at
$t=0$ that can be described in the spin-basis as
\begin{equation}
\hat{R}_{-\hat{y}}\left(\frac{\pi}{2}\right)\equiv\left(\begin{array}{cc}
R_{\uparrow\uparrow}^{(\pi/2)}\mathbf{1} & R_{\uparrow\downarrow}^{(\pi/2)}\mathbf{1}\\
R_{\downarrow\uparrow}^{(\pi/2)}\mathbf{1} & R_{\downarrow\downarrow}^{(\pi/2)}\mathbf{1}
\end{array}\right)=\frac{1}{\sqrt{2}}\left(\begin{array}{cc}
\mathbf{1} & \mathbf{1}\\
-\mathbf{1} & \mathbf{1}
\end{array}\right),
\end{equation}
where $\mathbf{1}$ represents the identity in the fermionic Hilbert
space. For simplicity, we denote $\hat{\mathcal{R}}\equiv\hat{R}_{-\hat{y}}\left(\pi/2\right)$
hereafter. The total time evolution is determined by the unitary transformation
\begin{equation}
\hat{\mathcal{U}}(t)=\hat{U}(t)\hat{\mathcal{R}},
\end{equation}
where
\begin{equation}
\hat{U}(t)=\left(\begin{array}{cc}
e^{-i\mathcal{\hat{H}}_{\uparrow}t} & 0\\
0 & e^{-i\mathcal{\hat{H}}_{\downarrow}t}
\end{array}\right)
\end{equation}
is the free time evolution operator in the spin basis representation
after the RF pulse. The final state of the system is thus given by
$\rho_{f}=\mathcal{U}\rho_{i}\mathcal{U}^{\dagger}$. Recall that
$S(t)=\langle\sigma_{-}\rangle$, we arrive at
\begin{equation}
S(t)={\rm Tr}\left(\sigma_{-}\rho_{f}\right)={\rm Tr}\left(e^{i\hat{\mathcal{H}}_{\downarrow}t}e^{-i\mathcal{\hat{H}}_{\uparrow}t}\rho_{{\rm FS}}\right)\label{eq:S(t)}
\end{equation}
that reduces to Eq. (\ref{eq:1DRamsey0T}) at zero temperature $k_{B}T^{\circ}=0$.

Since the complexity of the many-body Hamiltonians increases exponentially
with the number of particles in the system, an exact calculation of
$S(t)$ is usually inaccessible. However, in the case that $H_{\uparrow}$
and $H_{\downarrow}$ are both fermionic, bilinear many-body operators
as in Eq. (\ref{eq:Hamiltonian_idealfermion}), the overlap function
can reduce to a determinant in single-particle Hilbert space. This
approach, namely FDA, is based on a mathematical trace formula that
has been elegantly proved by Klich \cite{Klich2003Book}. (See Appendix
\ref{sec:ProofFDA} for details.) To proceed, we define $\hat{\mathcal{H}}_{\downarrow}\equiv\Gamma(h_{\downarrow})$
and $\hat{\mathcal{H}}_{\uparrow}\equiv\Gamma(h_{\uparrow})+\omega_{s}$.
Here, $\Gamma(h)\equiv\sum_{\mathbf{k},\mathbf{q}}h_{\mathbf{k}\mathbf{q}}c_{\mathbf{k}}^{\dagger}c_{\mathbf{q}}$
is a bilinear fermionic many-body Hamiltonian in the Fock space, and
$h_{\mathbf{k}\mathbf{q}}$ represents the matrix elements of the
corresponding operator in the single-particle Hilbert space. Applying
the Levitov\textquoteright s formula, Eq. (\ref{eq:FDA}), gives $S(t)=e^{-i\omega_{s}t}\tilde{S}(t)$
where

\begin{equation}
\tilde{S}(t)={\rm det}[(\mathbf{1}-\hat{n})+R(t)\hat{n}],
\end{equation}
and
\begin{equation}
R(t)=e^{ih_{\downarrow}t}e^{-ih_{\uparrow}t}.
\end{equation}
Correspondingly, the frequency domain spectrum can be obtained by
a Fourier transformation

\begin{equation}
A(\omega)=\frac{1}{\pi}\int_{0}^{\infty}e^{i\omega t}S(t)dt=\frac{1}{\pi}\int_{0}^{\infty}e^{i\tilde{\omega}t}\tilde{S}(t)dt.\label{eq:Aw}
\end{equation}
Hereafter, unless specified otherwise, we denote $\tilde{\omega}=\omega-\omega_{s}$
for any frequency variable $\omega$. As one can see, $\omega_{s}$
has a simple effect as shifting the frequency origin of a 1D spectrum.
Numerical calculations are carried out in a finite system confined
in a sphere of radius $R$. Keeping the density constant, we increase
$R$ towards infinity until numerical results are converged. Other
details of numerical calculations are described in the Appendix.

Figures \ref{fig:Ramsey1DFES} (a) and (c) show $\tilde{S}(t)$ for
attractive ($k_{F}a=-2$) and repulsive impurity interaction ($k_{F}a=+2$),
respectively. The zero-temperature (solid blue curves) asymptotic
behavior of the Ramsey response at $t\rightarrow\infty$ is given
by 
\begin{equation}
\begin{aligned}\tilde{S}(t)\simeq & Ce^{-i\Delta Et}\left(\frac{1}{iE_{F}t+0^{+}}\right)^{\alpha}\\
 & +C_{b}e^{-i\left(\Delta E-E_{F}+E_{b}\right)t}\left(\frac{1}{iE_{F}t+0^{+}}\right)^{\alpha_{b}},
\end{aligned}
\label{eq:Sfit_idealF}
\end{equation}
where $C$ and $C_{b}$ are both numerical constants independent with
respect to $k_{F}a$ and $C_{b}=0$ for $a<0$. Here, 
\begin{equation}
\alpha=\eta(E_{F})^{2}/\pi^{2}
\end{equation}
and 
\begin{equation}
\alpha_{b}=[1+\eta(E_{F})/\pi]^{2}
\end{equation}
are determined by the $s$-wave scattering phase shifts $\eta(E_{F})$
at Fermi energy. $E_{b}$ is the binding energy of the shallowest
bound state consisting of the impurity and a spin-up fermion for $a_{\uparrow}>0$
and $\Delta=0$. Furthermore, the change in energy can be understood
as a renormalization of the Fermi sea by impurity scattering and is
given by 
\begin{equation}
\Delta E=\sum_{\nu=1}^{N}(E_{\nu}-\tilde{E}_{\nu}),
\end{equation}
where $E_{\nu}$ and $\tilde{E}_{\nu}$ are the lowest $N$ eigenenergies
of $\hat{h}_{\downarrow}$ and $\hat{h}_{\uparrow}$, respecitively,
and the deeply bound states are excluded from $\tilde{E}_{\nu}$.
Here $N$ is the number of particles fixed by the chemical potential
$\mu$.

The corresponding spectral function ${\rm Re}A(\omega)$ is shown
in Fig. \ref{fig:Ramsey1DFES} (b) and (d). The asymptotic behaviors
of $\tilde{S}(t)$ translate into the threshold behaviors of spectra
function at zero temperature. For frequency $\tilde{\omega}\approx\Delta E$,
we have a singularity ${\rm Re}A(\omega)\propto\theta(\tilde{\omega}-\Delta E)|\tilde{\omega}-\Delta E|^{\alpha-1}$.
If the impurity interaction is repulsive, an additional singularity
shows up at $\tilde{\omega}\approx\omega_{b}=\Delta E-E_{F}+E_{b}$
as ${\rm Re}A(\omega)\propto\theta(\tilde{\omega}-\omega_{b})|\tilde{\omega}-\omega_{b}|^{\alpha_{b}-1}$.
These singularities are called FES and are closely related to the
polaron resonances (see Fig. \ref{fig:Ramsey1DMIBCS} for example):
the spectrum only shows one peak for attractive impurity interaction
and shows two peaks for repulsive interaction. We, therefore, name
the FESs in Fig. \ref{fig:Ramsey1DFES} (b) and (d) as attractive
or repulsive singularities, denoted by ``A'' and ``R'', respectively.
However, different than the polaron peaks that are Dirac delta functions
or Lorentzians, the FESs are power-law singularities, which is a manifestation
of OC: the quasiparticle resonances are rendered into power-law singularities
by the multiple particle-hole excitations near the surface of Fermi
sea. At a finite temperature, however, the thermal fluctuation leads
to an exponential decay $\tilde{S}(t)$ and Lorentzian-shape broadening
of singularities in ${\rm Re}A(\omega)$, which allows FDA calculations
to quantitatively predict the spectrum of mobile polaron if the thermal
fluctuation is comparable to the recoil energy \cite{Demler2016Science}.

\section{A BCS superfliuid as a background medium}

\begin{figure}
\includegraphics[width=0.98\columnwidth]{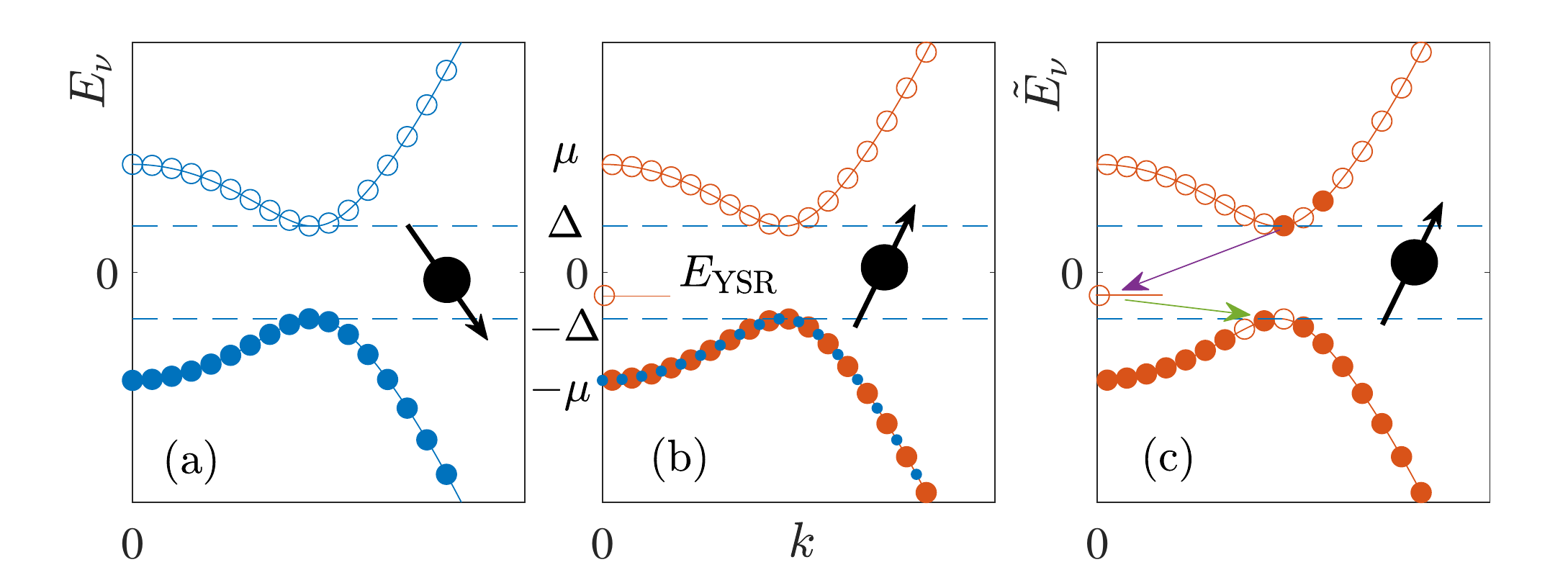}\caption{A sketch of the occupation and structure of the single-particle dispersion
spectrum of a two-component superfluid Fermi gas with a positive chemical
potential $\mu>0$ and the presence of a static impurity (black dot).
(a) shows the spectrum when the impurity is in the noninteracting
state (black arrow down) at zero temperature. When the impurity is
in the interacting polaron state (black arrow up), the spectrum is
shown in (b) at zero and (c) at finite temperature. Reprinted with
permission from \cite{JiaWang2022PRLshort}. \label{fig:SketchBCS}}
\end{figure}
\begin{figure}
\includegraphics[width=0.98\columnwidth]{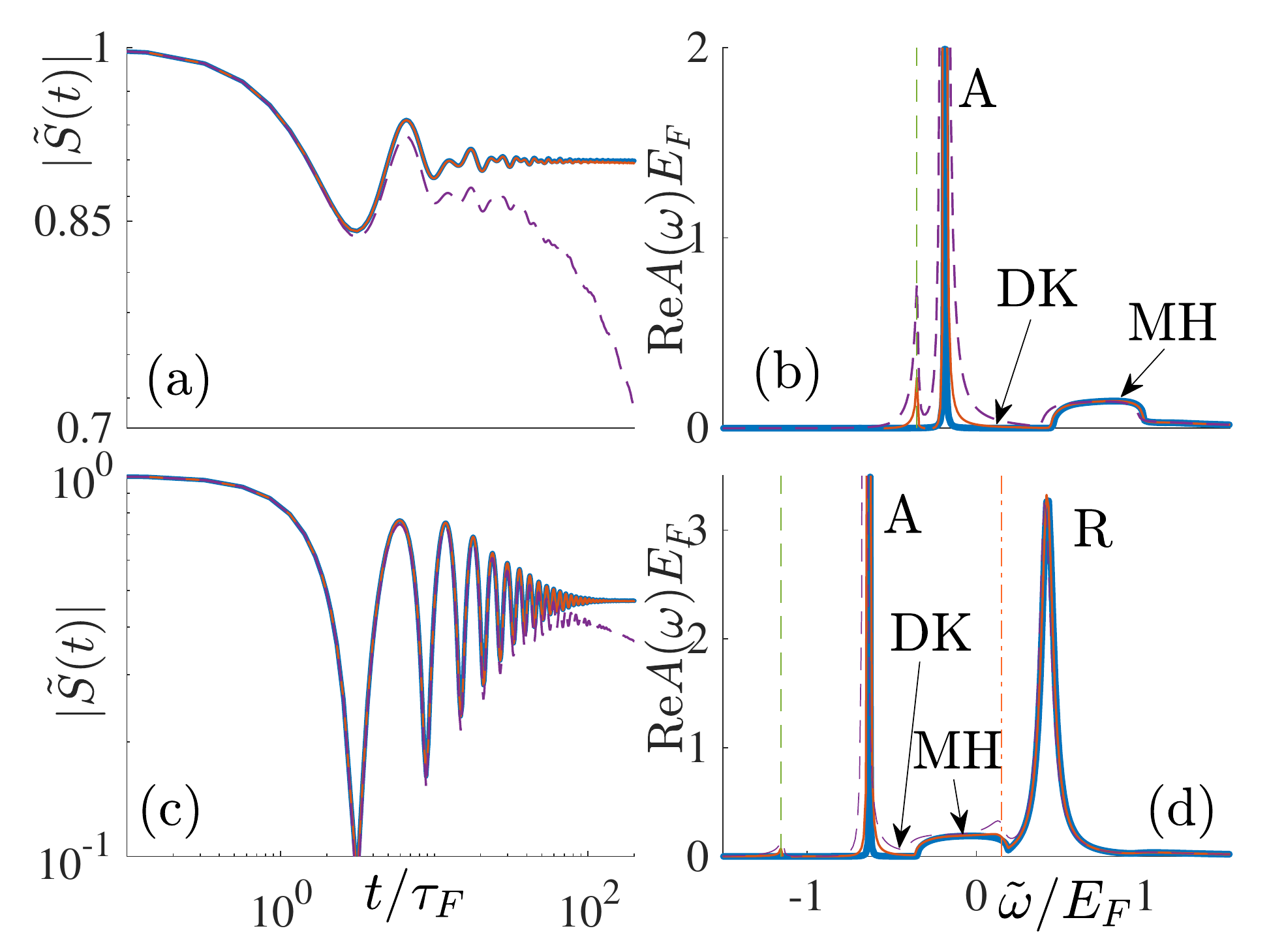}

\caption{1D Ramsey spectroscopy of magnetic impurity ($k_{F}a_{\Downarrow}=0$)
in a BCS superfluid at the BCS side ($k_{F}a_{F}=-2$) for (a) (b)
attractive impurity interaction $k_{F}a_{\Uparrow}=-2$ and (c) (d)
repulsive impurity interaction $k_{F}a_{\Uparrow}=2$. (a) and (c)
show the overlap functions $\tilde{S}(t)$. (b) and (d) show the spectral
functions ${\rm Re}A(\omega)$. Thick blue curves correspond to $k_{B}T^{\circ}=0$,
thin red solid curves, and purple dashed curves correspond to $k_{B}T^{\circ}=0.1E_{F}$
and $k_{B}T^{\circ}=0.15E_{F}$, respectively. A and R indicate the
attractive and repulsive polaron resonances, respectively. DK and
MH denote the dark continuum and molecule-hole continuum correspondingly.
The green dashed, and purple dash-dotted lines indicate the YSR features
$E_{{\rm RSR}}^{(-)}$ and $E_{{\rm RSR}}^{(+)}$, respectively. \label{fig:Ramsey1DMIBCS}}
\end{figure}

This section extends the FDA to a strongly correlated superfluid background
described by a standard BCS mean-field wavefunction \cite{JiaWang2022PRLshort,JiaWang2022PRAlong}.
The purpose is twofold. First, we aim to construct an exactly solvable
model for polaron with finite residue. This study shows that our system
is suitable for an exact approach --- an extended FDA. The presence
of a pairing gap can efficiently suppress multiple particle-hole excitations
and prevent Anderson's OC. Therefore, our model provides a benchmark
calculation of the polaron spectrum and rigorously examines all the
speculated polaron features. We name our system ``heavy crossover
polaron'' since the background Fermi gas can undergo a crossover
from a Bose-Einstein condensation (BEC) to a BCS superfluid. Second,
our prediction can be applied to investigate the background Fermi
superfluid excitations, a long-standing topic in ultracold atoms.
Polarons have already been realized in BEC and ideal Fermi gas experimentally,
where the physics of these weakly interacting background gas is well
understood. More recently, it has also been shown that polarons in
BEC with a synthetic spin-orbit-coupling can reveal the nature of
the background roton excitations \citep{Jia2019PRL}. Investigating
polaron physics in a strongly correlated Fermi superfluid at the BEC-BCS
crossover, namely crossover polaron, has also been proposed in several
pioneering works with approximated approaches \cite{Nishida2015PRL,Yi2015PRA,Pierce2019PRL,HuiHu2022PRA1,Bigue2022PRA}.
Our exact method in the heavy impurity limit allows us to apply the
polaron spectrum to measure the Fermi superfluid excitation features,
such as the pairing gap and sub-gap Yu-Shiba-Rusinov (YSR) bound states
\citep{Yu1965ActaPhysSin,Shiba1968ProgTheorPhys,Rusinov1969JETP,Vernier2011PRA,Jiang2011PRA}.

Our system consists of a localized impurity atom and a two-component
Fermi superfluid with equal mass $m_{\Uparrow}=m_{\Downarrow}=m$.
(Here, we use $|\Uparrow\rangle$ and $|\Downarrow\rangle$ to represent
the two internal states of the background fermionic atoms, in contrast
to the $|\uparrow\rangle$ and $|\downarrow\rangle$ for the impurity.)
The interaction between unlike atoms in the two-component Fermi gas
can be tuned by a broad Feshbach resonance and characterized by the
$s$-wave scattering length $a_{F}$. At low temperatures $T$, these
strongly interacting fermions undergo a crossover from a BEC to a
BCS superfluid, which can be described by the celebrated BCS theory
at a mean-field level. The full Hamiltonian can also be written in
the form of Eq. \ref{eq:HamiltonianForm}, where $\mathcal{\hat{H}}_{\uparrow}=\mathcal{\hat{H}}_{\downarrow}+\omega_{s}+\hat{V}$,

\begin{equation}
\hat{V}=\sum_{\sigma=\Uparrow,\Downarrow}\sum_{\mathbf{k},\mathbf{q}}\tilde{V}_{\sigma}(\mathbf{k}-\mathbf{q})c_{\mathbf{k}\sigma}^{\dagger}c_{\mathbf{q}\sigma},
\end{equation}
with $\tilde{V}_{\sigma}(\mathbf{k})$ being the Fourier transformation
of the potential between impurity and $\sigma$-component fermion
$V_{\sigma}(\mathbf{r})$, and
\begin{equation}
\mathcal{\hat{H}}_{\downarrow}=\hat{H}_{{\rm BCS}}\equiv K_{0}+\sum_{\mathbf{k}}\hat{\psi}_{\mathbf{k}}^{\dagger}\underline{h_{\downarrow}(\mathbf{k})}\hat{\psi}_{\mathbf{k}},\label{eq:Hiblinear}
\end{equation}
is the standard BCS Hamiltonian for noninteracting impurity. Here,
$\hat{\psi}_{\mathbf{k}}^{\dagger}\equiv(c_{\mathbf{k}\Uparrow}^{\dagger},c_{-\mathbf{k}\Downarrow})$
is the Nambu spinor representation, with $c_{\mathbf{k}\Uparrow}^{\dagger}$
($c_{\mathbf{k}\Downarrow}$) being the creation (annihilation) operator
for a $\sigma$-component fermion with momentum $\mathbf{k}$. $K_{0}\equiv-\mathcal{V}\Delta^{2}/g+\sum_{\mathbf{k}}(\epsilon_{\mathbf{k}}-\mu)$,
with $\mathcal{V}$ denoting the system volume and $\Delta$ being
the pairing gap parameter. $\epsilon_{\mathbf{k}}\equiv\hbar^{2}k^{2}/2m$
is the single-particle dispersion relation, and $\mu$ is the chemical
potential. We assume the populations of the two components are the
same and fixed by $\mu$. The bare coupling constant $g$ should be
renormalized by the $s$-wave scattering length $a_{F}$ between the
two components via $g^{-1}=m/4\pi a_{F}-\sum_{\mathbf{k}}^{\Lambda}1/2\epsilon_{\mathbf{k}}$,
where $\Lambda$ is an ultraviolet cut--off. $\underline{h_{\downarrow}(\mathbf{k})}$
can be regarded as a single-particle Hamiltonian $\hat{h}_{\downarrow}$
in momentum space and has a matrix form:

\begin{equation}
\underline{h_{\downarrow}(\mathbf{k})}=\left[\begin{array}{cc}
\xi_{\mathbf{k}} & \Delta\\
\Delta & -\xi_{\mathbf{k}}
\end{array}\right],\label{eq:hBCS}
\end{equation}
where $\xi_{\mathbf{k}}\equiv\epsilon_{\mathbf{k}}-\mu$. For a given
scattering length $a_{F}$ and temperature $T^{\circ}$, $\Delta$
and $\mu$ are determined by a set of the mean-field number and gap
equations \cite{Gurarie2006AnnPhys}.

To apply FDA, we need to express $\mathcal{H}_{\downarrow}$ and $\mathcal{H}_{\uparrow}$
in a bilinear form. It would be convenient to define $\hat{\psi}_{\mathbf{k}}^{\dagger}=(c_{\mathbf{k}\Uparrow}^{\dagger},c_{-\mathbf{k}\Downarrow})\equiv(c_{\mathbf{k}}^{\dagger},h_{\mathbf{k}}^{\dagger})$
and rewrite $\hat{V}$ as 
\begin{equation}
\hat{V}=\sum_{\mathbf{k\mathbf{q}}}\left[\tilde{V}_{\Uparrow}(\mathbf{k}-\mathbf{q})c_{\mathbf{k}}^{\dagger}c_{\mathbf{q}}-\tilde{V}_{\Downarrow}(\mathbf{q}-\mathbf{k})h_{\mathbf{k}}^{\dagger}h_{\mathbf{q}}\right]+\sum_{\mathbf{k}}\tilde{V}_{\Downarrow}(0),
\end{equation}
making the bilinear form apparent. We can also write the bilinear
form of $\mathcal{\hat{H}}_{\uparrow}$ explicitly as

\begin{equation}
\hat{\mathcal{H}}_{\uparrow}=K_{0}+\omega_{0}+\omega_{s}+\sum_{\mathbf{k\mathbf{q}}}\hat{\psi}_{\mathbf{k}}^{\dagger}\underline{h_{\uparrow}(\mathbf{k},\mathbf{q})}\hat{\psi}_{\mathbf{q}},\label{eq:Hfbilinear}
\end{equation}
where $\omega_{0}=\sum_{\mathbf{k}}\tilde{V}_{\Downarrow}(0)$ and

\begin{equation}
\underline{h_{\uparrow}(\mathbf{k},\mathbf{q})}=\underline{h_{\downarrow}(\mathbf{k})}\delta_{\mathbf{k}\mathbf{q}}+\left[\begin{array}{cc}
\tilde{V}_{\Uparrow}(\mathbf{k}-\mathbf{q}) & 0\\
0 & -\tilde{V}_{\Downarrow}(\mathbf{q}-\mathbf{k})
\end{array}\right]
\end{equation}
can be regarded as a single-particle Hamiltonian $\hat{h}_{\uparrow}$
in momentum space and in a matrix form. We can see that, $\hat{h}_{\uparrow}$
and $\hat{h}_{\downarrow}$ are the single-particle representative
of $\hat{\mathcal{H}}_{\uparrow}$ and $\hat{\mathcal{H}}_{\downarrow}$
up to some constants, respectively.

Diagonalizing $\underline{h_{\downarrow}(\mathbf{k})}$ gives the
well-known BCS dispersion relation $E_{\nu}=\pm\mathcal{E}_{\mathbf{k}}=\pm\sqrt{\xi_{\mathbf{k}}^{2}+\Delta^{2}}$,
where $\nu\equiv\{\pm,\mathbf{k}\}$ is a collective index. As sketched
in Fig. \ref{fig:SketchBCS} (a), this spectrum consists of positive
and negative branches separated by an energy gap. Since we prepare
the impurity initially in the noninteracting state, the atoms occupy
the eigenstates of $\underline{h_{\downarrow}(\mathbf{k})}$ with
a Fermi distribution $f(E_{\nu})=1/\left(e^{-E_{\nu}/k_{B}T}+1\right)$.
At zero temperature, the many-body ground state can be regarded as
a fully filled Fermi sea of the lower branch and a completely empty
Fermi sea of the upper branch. When the impurity interaction is on,
eigenvalues $\tilde{E}_{\nu}$ of $\underline{h_{\uparrow}(\mathbf{k})}$
still consists of two branches separated by the same gap, with each
individual energy level shifted, as shown in Fig. \ref{fig:SketchBCS}
(b). Moreover, when the impurity scattering is magnetic ($a_{\Uparrow}\ne a_{\Downarrow})$,
a sub-gap YSR bound state exists \cite{Yu1965ActaPhysSin,Shiba1968ProgTheorPhys,Rusinov1969JETP,Vernier2011PRA,Jiang2011PRA}.

It is worth noting that, in the many-body Hamiltonian $\hat{\mathcal{H}}_{\uparrow}$,
we have assumed that the pairing order parameter $\Delta$ remains
unchanged by introducing the interaction potential $V_{\sigma}(\mathbf{r})$.
For a non-magnetic potential ($V_{\Uparrow}=V_{\Downarrow}$) that
respects time-reversal symmetry, this is a reasonable assumption,
according to Anderson's theorem \citep{Balatsky2006RMP}. For a magnetic
potential ($V_{\Uparrow}\neq V_{\Downarrow}$), the local pairing
gap near the impurity will be affected, as indicated by the presence
of the YSR bound state. We will follow the typical non-self-consistent
treatment of the magnetic potential in condensed matter physics \citep{Balatsky2006RMP,Yu1965ActaPhysSin}
and assume a constant pairing gap as the first approximation for simplicity.

Inserting the bilinear forms of Hamiltonian into the expression of
overlap function in Eq. (\ref{eq:S(t)}) and applying FDA gives $S(t)=e^{-i\omega_{s}t}\tilde{S}(t)$
where
\begin{equation}
\tilde{S}(t)=e^{-i\omega_{0}t}{\rm det}[1-\hat{n}+e^{i\hat{h}_{\downarrow}t}e^{-i\hat{h}_{\uparrow}t}\hat{n}],
\end{equation}
with $\hat{n}$ is the occupation number operator. The corresponding
spectral function in the frequency domain is given by Eq. (\ref{eq:Aw}).

Figure \ref{fig:Ramsey1DMIBCS} shows numerical results for a magnetic
impurity ($k_{F}a_{\Downarrow}=0$) immersed in the background BCS
superfluid at the BCS side ($k_{F}a_{F}=-2$). In sharp contrast to
the noninteracting Fermi gases, for cases with a nonzero pairing gap,
the asymptotic behavior in the long-time limit shows that $|\tilde{S}(t\rightarrow\infty)|\propto t^{0}$
approaches a constant. These asymptotic constants are larger for larger
$\Delta$. Further details can be obtained by an asymptotic form that
fits our numerical calculations perfectly
\begin{equation}
\tilde{S}(t)\simeq D_{a}e^{-iE_{a}t}+D_{r}e^{-iE_{r}t},\label{eq:Sfit_Mag}
\end{equation}
where $D_{r}=0$ for $a_{\uparrow}<0$. We obtain $D_{a}$, $D_{r}$,
$E_{a}$ and $E_{r}$ from fitting and find that $E_{r}={\rm Re}E_{r}+i{\rm Im}E_{r}$
is, in general, complex. In contrast, $E_{a}=\sum_{E_{\nu}<0}(E_{\nu}-\tilde{E}_{\nu})$
(where $\tilde{E}_{\nu}$ excludes the two-body deeply bound states)
is purely real and can be explained as a renormalization of the filled
Fermi sea.

The long-time asymptotic behavior of $S(t)$ manifests itself as some
characterized lineshape in the spectral function 
\begin{equation}
A(\omega)\propto\begin{cases}
Z_{a}\delta(\tilde{\omega}-E_{a}) & \tilde{\omega}\approx E_{a}\\
Z_{r}\frac{\left|{\rm Im}E_{r}\right|/\pi}{(\tilde{\omega}-{\rm Re}E_{r})^{2}+({\rm Im}E_{r})^{2}} & \tilde{\omega}\approx{\rm Re}E_{r}
\end{cases},
\end{equation}
i.e., a $\delta$-function around $E_{a}$ and a Lorentzian around
${\rm Re}E_{r}$. The existence of the $\delta$-function peak unambiguously
confirms the existence of a well-defined quasiparticle -- the attractive
polaron with energy $E_{a}$. The Lorentzian, on the other hand, can
be recognized as a repulsive polaron with finite width and hence finite
lifetime. Here, $Z_{a}=|D_{a}|$ and $Z_{r}=|D_{r}|$ are the residues
of attractive and repulsive polaron, correspondingly. Numerically,
we find that $Z_{a}\propto(\Delta/E_{F})^{\alpha_{a}}$ and $Z_{r}\propto(\Delta/E_{F})^{\alpha_{r}}$
at small $\Delta$. The existence of finite residue of polarons indicates
that the pairing gap suppresses multiple particle-hole excitations
and prevents OC, which eventually leads to the survival of well-defined
polarons.

We also find that the attractive polaron separates from a molecule-hole
continuum (denoted as MH in Fig. \ref{fig:Ramsey1DMIBCS}) by a region
of anomalously low spectral weight, namely the ``dark continuum''
(denoted as DK in Fig. \ref{fig:Ramsey1DMIBCS}). The existence of
a dark continuum has been previously observed in spectra of other
polaron systems. However, most of these studies apply various approximations,
and only recently, a diagrammatic Monte Carlo study proves the dark
continuum is indeed physical \citep{Goulko2016PRA}. Here, our FDA
calculation of the heavy crossover polaron spectrum gives exact proof
of the dark continuum. By comparing Fig. \ref{fig:Ramsey1DMIBCS}
and Fig. \ref{fig:Ramsey1DFES}, we expect that the dark continuum
vanishes in the $\Delta\rightarrow0$ limit and the attractive polaron
merges into the molecule-hole continuum, forming a power-law singularity
with the ``wing''. Similar behavior also can be observed for the
repulsive polaron, where the associated molecule-hole continuum is
much less significant and cannot be visually seen in Fig. \ref{fig:Ramsey1DMIBCS}.

Finite-temperature results are indicated by the thin red solid (purple
dashed) curves for $k_{B}T=0.1E_{F}\ (0.15E_{F})$ in Fig. \ref{fig:Ramsey1DMIBCS}.
Some surprising features show up, other than the expected thermal
broadening. An enhancement of spectral weight appears sharply at the
energy $E_{{\rm YSR}}^{(-)}=E_{a}-(\Delta-E_{{\rm YSR}})$ below the
attractive polaron. This spectral feature corresponds to the decay
process highlighted by the purple arrow in Fig. (c), where an additional
particle initially excited to the upper Fermi sea by thermal fluctuation
is driven to the YSR state. For the case of $k_{F}a_{\uparrow}>0$,
a feature associated with the repulsive polaron appears at $E_{{\rm YSR}}^{(+)}={\rm Re}(E_{r})-(E_{{\rm YSR}}+\Delta)$,
as indicated by the green arrow in Fig. \ref{fig:SketchBCS} (c):
an additional particle decays from the YSR state to the lower Fermi
sea. The polaron spectrum can be applied to measure the superfluid
gap $\Delta$ and $E_{{\rm YSR}}$. In particular, we notice, on the
positive side $a_{\uparrow}>0$, if $E_{a}$, ${\rm Re}(E_{r})$,
$E_{{\rm YSR}}^{(-)}$ and $E_{{\rm YSR}}^{(+)}$ can all be measured
accurately, we have $2\Delta=E_{a}+{\rm Re}(E_{r})-E_{{\rm YSR}}^{(-)}-E_{{\rm YSR}}^{(+)}$
that does not depend on $E_{{\rm YSR}}$. Since this formula only
relies on the existence of the gap and a mid-gap state, we anticipate
it can be used to measure $\Delta$ accurately for a Fermi superfluid
that can not be quantitatively described by the BCS theory.

\section{Multidimension spectroscopy}

\begin{figure}
\includegraphics[width=0.98\columnwidth]{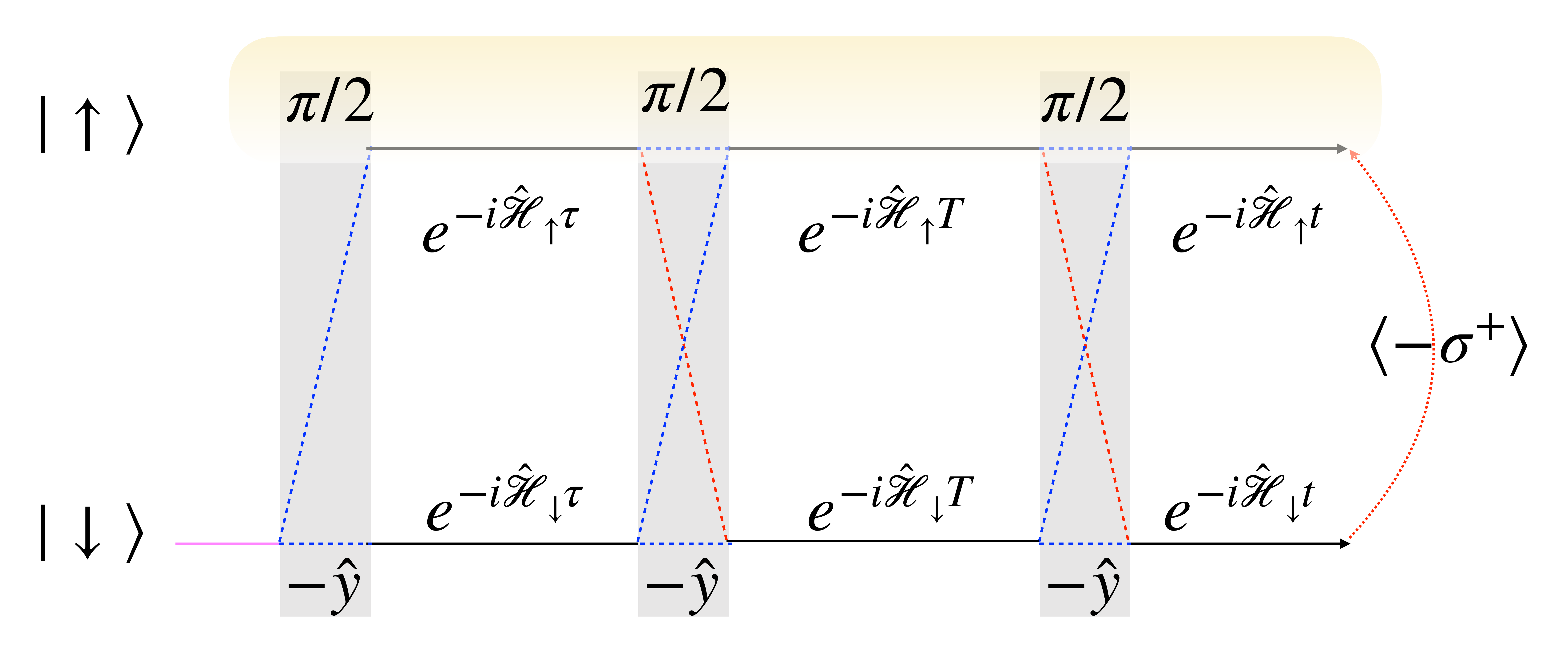}

\caption{A sketch of EXSY (EXchange SpectroscopY) pulses scheme. \label{fig:SketchMD}}
\end{figure}
\begin{figure*}
\includegraphics[width=0.98\columnwidth]{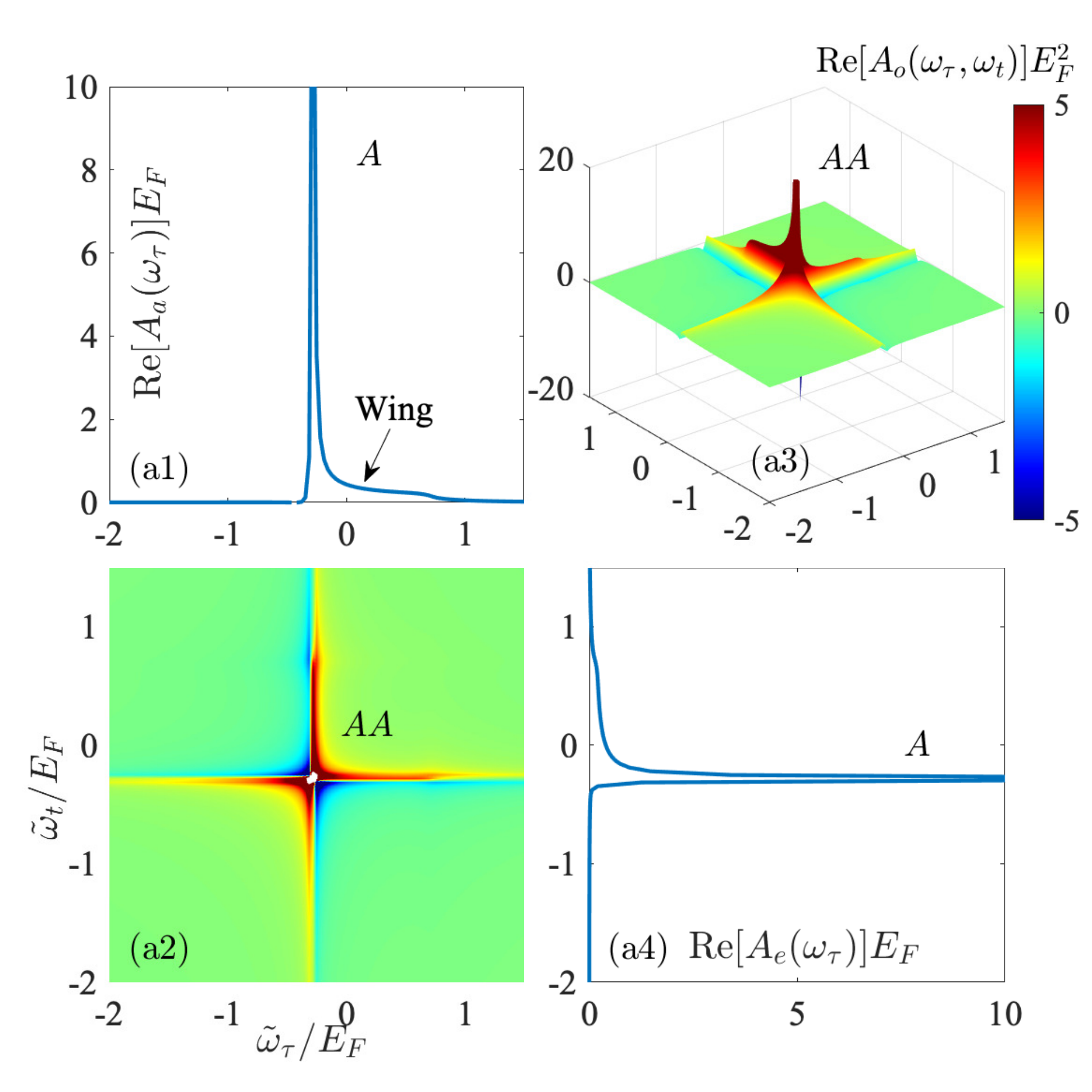}\includegraphics[width=0.98\columnwidth]{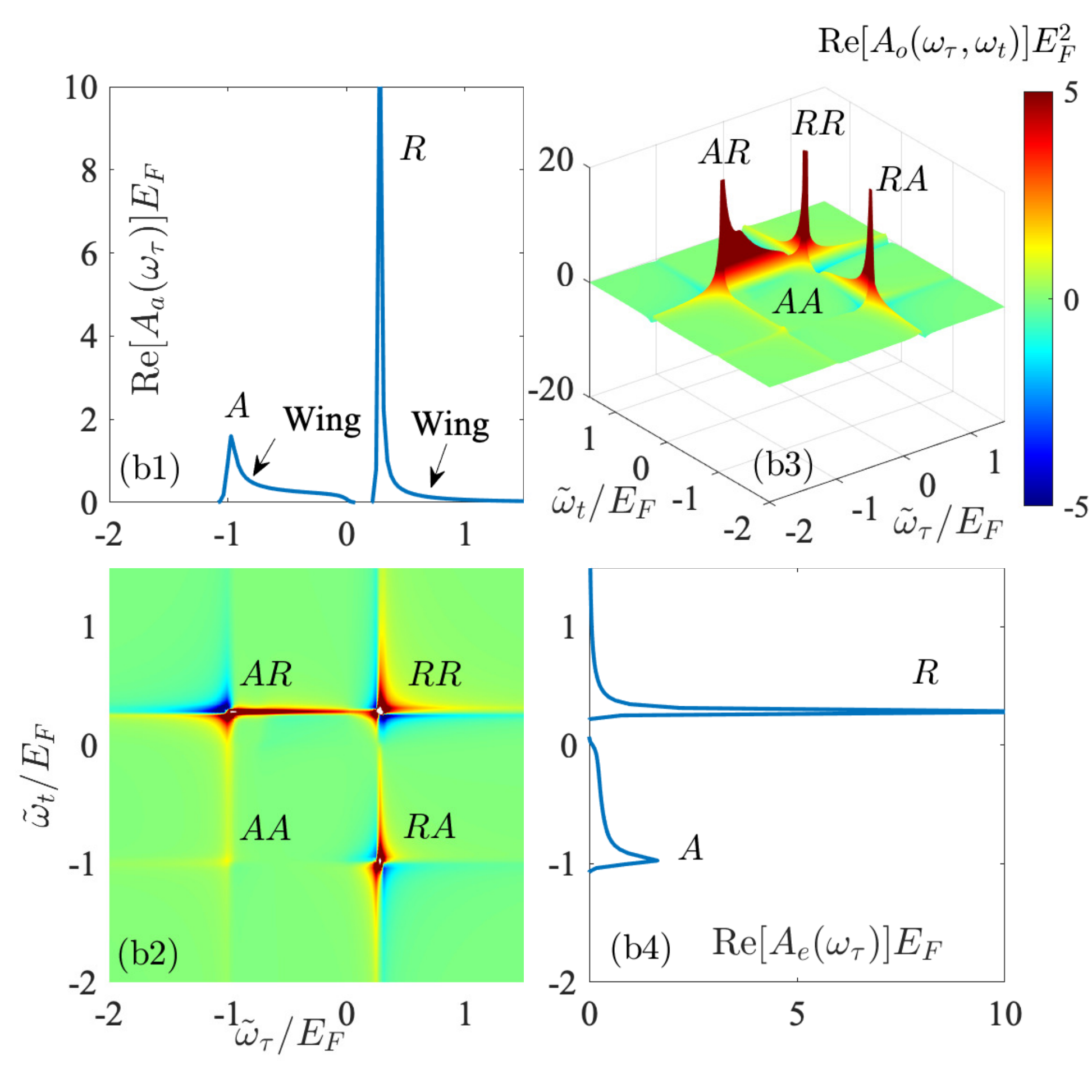}\caption{(a1) and (a4) shows the 1D absorption spectrum for attractive interaction
$k_{F}a=-0.05$ and finite temperature $k_{B}T^{\circ}=0.03E_{F}$.
The absorption singularity is denoted as $A$. (a2), and (a3) shows
the contour, and 3D landscape of the 2D spin-echo spectrum ${\rm Re}[A_{o}(\omega_{\tau},\omega_{t})]$,
where the diagonal peak is denoted as $AA$. (b1)-(b4) are the same
as (a1)-(a4), correspondingly, but for repulsive interaction $k_{F}a=0.5$.
There are two singularities in (b1), the absorption spectrum, namely
repulsive and attractive singularities, which are denoted as $R$
and $A$. The corresponding diagonal peaks in (b2) and (b3) are denoted
as $AA$ and $RR$, while the off-diagonal cross-peaks are denoted
as $AR$ and $RA$. Reprinted with permission from \cite{JiaWang2022arXiv1}
\label{fig:Aw}}
\end{figure*}

In this section, we present another new extension of the FDA in the
calculation of multidimensional (MD) Ramsey spectroscopy \cite{JiaWang2022arXiv1}.
Conventional Ramsey spectroscopy, such as the ones studied in previous
sections, is called 1D since it shows the signal as a function of
only one variable: the frequency of the single applied RF pulse or
the time between the RF pulse and measurement. Here, we investigate
a scenario where multiple RF pulses manipulate the impurity at several
different times. The observed signal's dependency on the time intervals
between pulses or the corresponding Fourier transformation is called
MD Ramsey spectroscopy.

Pushing 1D Ramsey spectroscopy to MD shares the same spirit as the
widely successful MD nuclear magnetic resonance (NMR) and optical
MD coherent spectroscopy (MDCS). MD spectroscopy not only improves
the resolution and overcomes spectral congestion but also carries
rich information on the correlations between resonance peaks and provides
insights into physics that 1D spectroscopy cannot access. For example,
in a 2D NMR spectroscopy, the peaks on the diagonal map the resonances
in 1D spectroscopy; however, only coupled spins give rise to off-diagonal
cross-peaks between corresponding resonances. The cross peaks are
thus the signature of correlations between resonances, which the 1D
spectrum cannot distinguish. In our system, the correlations in MD
Ramsey spectroscopy are induced by the coupling between the spin and
the background Fermi gas, a genuine many-body environment, and hence
called many-body correlations.

We consider the same system described in Sec. \ref{subsec:FDA}, a
localized impurity immersed in a noninteracting Fermi gas but manipulated
by multiple RF pulses. One example of a three-pulse scheme is shown
in Fig. \ref{fig:SketchMD} (a), which is similar to one of the most
common 2D NMR pulse sequences, namely EXSY (EXchange SpectroscopY).
EXSY essentially measures the four-wave mixing response of our system.
The time evolution is thus given by the unitary transformation
\begin{equation}
\mathcal{U}(t,T,\tau)=U(t)\mathcal{R}U(T)\mathcal{R}U(\tau)\mathcal{R}.
\end{equation}
We define the MD responses as
\begin{equation}
S(\tau,T,t)=-{\rm Tr}(\sigma_{+}\rho_{f}),
\end{equation}
where the choice of $\sigma_{+}$ and additional $-$1 prefactor are
for conventions so that $S(\tau,T=0,t=0)$ is equivalent to the 1D
overlap function $S(\tau)$. Notice that we have the relation $\mathcal{R}^{-1}\mathcal{R}^{-1}\mathcal{\mathcal{\sigma_{-}\mathcal{R\mathcal{R}}=-\sigma_{+}}}$.

The multidimensional response $S(\tau,T,t)$ can be written as a summation
of sixteen contributions

\begin{equation}
S(\tau,T,t)=\sum_{i=1}^{16}S_{i}(\tau,T,t)\equiv\frac{1}{4}\sum_{i=1}^{16}\mathrm{Tr}[I_{i}(\tau,T,t)\rho_{{\rm FS}}],\label{eq:StauTt}
\end{equation}
where $I_{i}(\tau,T,t)$ are named pathways. These pathways are written
as a direct product of six free-evolution operators $e^{-i\mathcal{\hat{H}^{\prime}}t^{\prime}}$
or their complex conjugates, such as Eqs. (\ref{eq:path1}) and (\ref{eq:path2}).
Here, $\hat{\mathcal{H}}^{\prime}$ can be $\hat{\mathcal{H}}_{\uparrow}$
or $\hat{\mathcal{H}}_{\downarrow}$ and $t^{\prime}$ can be $\tau$,
$T$, or $t$. The expressions of pathways are recognized to be similar
to the optical paths in an interferometer as sketched in Fig. (\ref{fig:SketchMD}),
where the free evolution-operator is illustrated by the solid black
lines, the dashed lines in the grey beam splitter correspond to the
matrix elements $R_{\sigma\sigma^{\prime}}^{(\pi/2)}$ of the rotating
operator in the spin basis. The measurement operator $\sigma_{+}$
fixes the middle two terms that depend on $t$ as $...e^{i\hat{\mathcal{H}}_{\uparrow}t}e^{-i\mathcal{\hat{H}}_{\downarrow}t}...$,
and the remaining operators each have two possibilities, leading to
$2\times4=16$ possible combinations.

A summation of the contributions of all sixteen pathways gives the
total response $S(t,T,\tau)$, and the spectrum in the frequency domain
can be obtained via a double Fourier transformation
\begin{equation}
A(\omega_{\tau},T,\omega_{t})=\frac{1}{\pi^{2}}\int_{0}^{\infty}\int_{0}^{\infty}dtd\tau e^{i\omega_{\tau}\tau}S(\tau,T,t)e^{-i\omega_{t}t},\label{eq:Aw2D}
\end{equation}
where $\omega_{t}$ and $\omega_{\tau}$ are interpreted as an absorption
and emission frequency, respectively. On the other hand, the dependence
of $A(\omega_{\tau},T,\omega_{t})$ on the mixing time $T$ can reveal
the many-body coherent and incoherent dynamics \cite{Tempelaa2019NC}.
The physical process underlying $A(\omega_{\tau},T,\omega_{t})$ can
be interpreted as an inequilibrium dynamical evolution: the system
first gets excited by absorbing a photon with frequency $\omega_{\tau}$,
after a period of mixing time $T$, and then emits a photon with frequency
$\omega_{t}$. We notice that $A(\omega_{\tau},T,\omega_{t})=\sum_{i=1}^{16}A_{i}(\omega_{\tau},T,\omega_{t})$
can also be expressed as a summation of sixteen pathways, where the
expression of each pathway is given by Eq. (\ref{eq:Aw2D}), with
$A$ and $S$ replaced by $A_{i}$ and $S_{i}$, respectively.

We can take the rotating wave approximation and consider only two
dominant pathways (with details given by \cite{JiaWang2022arXiv1})
\begin{equation}
I_{1}(\tau,T,t)=e^{i\hat{\mathcal{H}}_{\downarrow}\tau}e^{i\hat{\mathcal{H}}_{\uparrow}T}e^{i\hat{\mathcal{H}}_{\uparrow}t}e^{-i\hat{\mathcal{H}}_{\downarrow}t}e^{-i\hat{\mathcal{H}}_{\uparrow}T}e^{-i\hat{\mathcal{H}}_{\uparrow}\tau},\label{eq:path1}
\end{equation}
\begin{equation}
I_{2}(\tau,T,t)=e^{i\hat{\mathcal{H}}_{\downarrow}\tau}e^{i\hat{\mathcal{H}}_{\downarrow}T}e^{i\hat{\mathcal{H}}_{\uparrow}t}e^{-i\hat{\mathcal{H}}_{\downarrow}t}e^{-i\hat{\mathcal{H}}_{\downarrow}T}e^{-i\hat{\mathcal{H}}_{\uparrow}\tau}.\label{eq:path2}
\end{equation}
It should be notice that the expression of $I_{1}(\tau,T,t)$ and
$I_{2}(\tau,T,t)$ are similar to those correspond to the excited
state emision (ESE) and ground state breaching (GSB) in the rephasing
2D coherent spectra \cite{HuiHu2022arXiv}.

The contribution of each pathway, $S_{i}(\tau,T,t)$, can be calculated
exactly via FDA. To proceed, we define $\mathcal{H}_{\downarrow}\equiv\Gamma(h_{\downarrow})$
and $\mathcal{H}_{\uparrow}\equiv\Gamma(h_{\uparrow})+\omega_{s}$.
Here $\Gamma(h)\equiv\sum_{\mathbf{k},\mathbf{q}}h_{\mathbf{k}\mathbf{q}}c_{\mathbf{k}}^{\dagger}c_{\mathbf{q}}$
is a bilinear fermionic many-body Hamiltonian in the Fock space, and
$h_{\mathbf{k}\mathbf{q}}$ represents the matrix elements of the
corresponding operator in the single-particle Hilbert space. These
matrix elements are explicitly given by $(h_{\downarrow})_{\mathbf{k}\mathbf{q}}=\epsilon_{\mathbf{k}}\delta_{\mathbf{k}\mathbf{q}}$
and $(h_{\uparrow})_{\mathbf{k}\mathbf{q}}=\epsilon_{\mathbf{k}}\delta_{\mathbf{k}\mathbf{q}}+\tilde{V}(\mathbf{k}-\mathbf{q})$.
With these definitions, we can rewrite
\begin{equation}
S_{i}(\tau,T,t)=\frac{1}{4}\tilde{S}_{i}(\tau,T,t)e^{-i\omega_{s}f_{i}(t,T,\tau)},\label{eq:Si}
\end{equation}
where $e^{-i\omega_{s}f_{i}(t,T,\tau)}$ gives a simple phase and
$\tilde{S}_{i}(\tau,T,t)$ is a product of the exponentials of the
bilinear fermionic operator, both of which can be calculated exactly.
For example, we have $S_{1}(\tau,T,t)=\tilde{S}_{1}(\tau,T,t)e^{i\omega_{s}t}e^{-i\omega_{s}\tau}/4$,
where
\begin{equation}
\begin{aligned}\tilde{S}_{1}(\tau,T,t)= & {\rm Tr}[e^{i\Gamma(h_{\downarrow})\tau}e^{i\Gamma(h_{\uparrow})T}e^{i\Gamma(h_{\uparrow})t}\times\\
 & e^{-i\Gamma(h_{\downarrow})t}e^{-i\Gamma(h_{\uparrow})T}e^{-i\Gamma(h_{\uparrow})\tau}\rho_{{\rm FS}}]
\end{aligned}
.
\end{equation}
Applying Levitov\textquoteright s formula \cite{Klich2003Book,JiaWang2022PRLshort,JiaWang2022PRAlong}
gives

\begin{equation}
\tilde{S}_{1}(\tau,T,t)={\rm det}[(1-\hat{n})+R_{1}(\tau,T,t)\hat{n}],
\end{equation}
with
\begin{equation}
R_{1}(\tau,T,t)=e^{ih_{\downarrow}\tau}e^{ih_{\uparrow}T}e^{ih_{\uparrow}t}e^{-ih_{\downarrow}t}e^{-ih_{\uparrow}T}e^{-ih_{\uparrow}\tau},
\end{equation}
and $\hat{n}=n_{\mathbf{k}}\delta_{\mathbf{k}\mathbf{k}'}$, where
$n_{\mathbf{k}}=1/(e^{\epsilon_{\mathbf{k}}/k_{B}T^{\circ}}+1)$ denotes
the single-particle occupation number operator.

The 2D spectrum $A_{o}(\omega_{\tau},\omega_{t})\equiv A(\omega_{\tau},T=0,\omega_{t})$
in Figs. \ref{fig:Aw} (a2) and (a3) shows a double dispersion lineshape
commonly found in 2D NMR around $(\tilde{\omega}_{\tau},\tilde{\omega}_{t})\approx(\tilde{\omega}_{A-},\tilde{\omega}_{A-})$,
which is called a diagonal peak denoted as $AA$. For attractive interaction
$k_{F}a=-0.5$, the attractive singularity appears at $\tilde{\omega}_{A-}\approx-0.28E_{F}$
in the absorption spectrum. We have numerically verified that the
integration of 2D spectroscopy over emission frequency $\omega_{t}$
gives the 1D absorption spectrum $A_{a}(\omega_{\tau})$ (not shown
here). Interestingly, we can observe that there is no diagonal spectral
weight corresponding to the wing. Rather, the spectral weight on the
off-diagonal $A_{o}(\omega_{\tau},\omega_{t}\approx\omega_{A-})$
and $A_{o}(\omega_{\tau}\approx\omega_{A-},\omega_{t})$ is significant
and resembles the lineshape of the wing. This is a non-trivial manifestation
of OC in the 2D spectroscopy: the inhomogeneous and homogeneous lineshape
does not have the OC characteristic, i.e., a power-law singularity
and a broad lineshape that resembles the wings in the 1D spectroscopy
\cite{Demler2012PRX}. Here, the inhomogeneous and homogeneous lineshape
refer to the lineshape near a singularity along the diagonal or the
direction perpendicular to the diagonal. As we can observe, the widths
of the singularity are much sharper along these two directions, which
might help experimental identification of the singularity, especially
at finite temperatures. The homogeneous and inhomogeneous broadenings
in MD spectroscopy also have their own experimental significance,
similar to their NMR or optical counterpart. In a realistic experiment,
the ensemble average of the impurity signal can give rise to a further
inhomogeneous broadening induced by the disorder of the local environment
(such as spatial magnetic field fluctuation). However, these disorders
are usually non-correlated and would not introduce homogeneous broadening
\cite{Cundiff2015JApplP,XiaoqinLi2016NL,XiaoqinLi2017NC}.

For repulsive interaction $k_{F}a=0.5$, there are two singularities,
the attractive and repulsive singularities, in the 1D absorption spectrum.
These singularities appear at $\tilde{\omega}_{A+}\approx-0.98E_{F}$
and $\tilde{\omega}_{R+}\approx0.28E_{F}$ in Figs. \ref{fig:Aw}
(b1) and (b4). As shown in Fig. \ref{fig:Aw} (b2) and (b3), there
are two diagonal peaks, $AA$ and $RR$, in the 2D spectroscopy that
mirror the attractive and repulsive singularities. In addition, there
are also two significant cross-peaks denoted as $AR$ and $RA$. The
physical interpretations of cross peaks are similar to those observed
in 2D NMR, where strong cross-peaks between the two spin resonances
indicate strong coupling between the two spins. In our system, the
correlation between attractive and repulsive singularities is induced
by the coupling between spin and the background Fermi gas, a many-body
environment, which is named a many-body quantum correlation. The strong
off-diagonal peaks, therefore, indicate a strong many-body quantum
correlation between the attractive and repulsive singularity induced
by the many-body environment. As far as we know, this is the first
prediction of many-body correlations between Fermi edge singularities
in cold atom systems. If the impurity has a finite mass or the background
Fermi gas is replaced by a superfluid with an excitation gap, we expect
these cross-peaks would remain and represent the correlations between
attractive and repulsive polarons \cite{JiaWang2022PRLshort,JiaWang2022PRAlong}.
The method reviewed here can also be straightforwardly applied to
calculate the coherent and relaxation dynamics of the system in terms
of the mixing-time $T$ dependence of the MD Ramsey spectroscopy \cite{JiaWang2022arXiv1}.
We also remark here that a calculation of the MD Ramsey spectroscopy
for a finite mass impurity has recently been developed using a Chevy's
ansatz approximation method \cite{JiaWang2022arXiv2}.
\begin{acknowledgments}
We are grateful to Hui Hu and Xia-Ji Liu for their insightful discussions
and critical reading of the manuscript. We also thank Jesper Levinsen
and Meera Parish for stimulating discussions.
\end{acknowledgments}

\appendix

\section{Klich's proof of a trace formula\label{sec:ProofFDA}}

One of the key equations in the functional determinant approach formalism
is a trace formula
\begin{equation}
{\rm Tr}\left[e^{\Gamma(A_{1})}e^{\Gamma(A_{2})}...e^{\Gamma(A_{N})}\right]={\rm det}\left(\mathbf{1}-\xi e^{A_{1}}e^{A_{2}}...e^{A_{N}}\right)^{-\xi},\label{eq:TraceFormula}
\end{equation}
where $\mathbf{1}$ is an identity matrix of the dimension of single-particle
Hilbert space, $\xi=1$ for bosons and $\xi=-1$ for fermions. Here
the many-body Fock space operator 
\begin{equation}
\Gamma(A_{n})=\sum_{ij}\langle i|A_{n}|j\rangle a_{i}^{\dagger}a_{j}\label{eq:FockOperator}
\end{equation}
is the second quantized version of a single particle operator $A_{n}$,
and $n\in\{1,2,...,N\}$ are integer subscripts. In contrast, $A_{n}$
is defined as an operator on the single particle Hilbert space, with
matrix elements $\langle i|A_{n}|j\rangle$, where $|i\rangle$ are
single-particle basis corresponding to the creation operator $a_{i}^{\dagger}$.
Here, we included the proof for completeness, mainly following Klich's
elegant proof \cite{Klich2003Book}.

First, we prove for a single operator, ${\rm Tr}\left[e^{\Gamma(A_{1})}\right]={\rm det}\left(\mathbf{1}-\xi e^{A_{1}}\right)^{-\xi}$.
We recall that any matrix $A_{1}$ can be written in a basis (corresponding
to creation operator $b_{i}^{\dagger}$) which it is of the form $D+K$,
where $D$ is a diagonal matrix with elements $D_{\nu\nu}\equiv\lambda_{\nu}$
known as eigenvalues of the matrix and $K$ is an upper triangular.
Since the upper trangular $K$ does not contribute to the trace, we
have

\begin{equation}
{\rm Tr}\left[e^{\Gamma(A_{1})}\right]={\rm Tr}\left[e^{\Gamma(D)}\right]={\rm Tr}\left[\prod_{\nu=1}^{K}e^{\lambda_{\nu}b_{\nu}^{\dagger}b_{\nu}}\right].
\end{equation}
Notice that the trace is taking over the Fock space basis $|\vec{\alpha}\rangle\equiv|\alpha_{1}\alpha_{2}...\alpha_{K}\rangle$
with $\alpha_{\nu}$ being the occupation number of the single-particle
basis $|\nu\rangle$ corresponding to $b_{\nu}^{\dagger}$. For fermions,
$\vec{\alpha}$ are vectors of zeros and ones and for bosons vectors
with integer coefficients. In such occupation number representation,
the trace can be expressed as
\begin{equation}
{\rm Tr}\left[\prod_{\nu=1}^{K}e^{\lambda_{\nu}b_{\nu}^{\dagger}b_{\nu}}\right]=\prod_{\nu=1}^{K}\sum_{\alpha_{\nu}}e^{\lambda_{\nu}\alpha_{\nu}},
\end{equation}
where

\begin{equation}
\sum_{\alpha_{\nu}}e^{\lambda_{\nu}\alpha_{\nu}}=(1-\xi e^{\lambda_{\nu}})^{-\xi}=\begin{cases}
1+e^{\lambda_{\nu}},\ ,\xi=-1 & {\rm Fermion},\\
1/(1-e^{\lambda_{\nu}}),\ \xi=1 & {\rm Boson}.
\end{cases}
\end{equation}
The fact that $\lambda_{\nu}$ are eigenvalues of $A_{1}$, which
implies $(1-\xi e^{\lambda_{\nu}})^{-\xi}$ are eigenvalues of $(1-\xi e^{A_{1}})^{-\xi}$,
leads to the products of eigenvalues $\prod_{\nu=1}^{K}(1-\xi e^{\lambda_{\nu}})^{-\xi}={\rm det}(\mathbf{1}-\xi e^{A_{1}})^{-\xi}$.
Consequently, we prove

\begin{equation}
{\rm Tr}\left[e^{\Gamma(A_{1})}\right]={\rm det}\left(\mathbf{1}-\xi e^{A_{1}}\right)^{-\xi}
\end{equation}
as promised. We remark that this formula does not depend on the single-state
basis, and an intuitive way of understanding this formula can be thinking
of ${\rm Tr}\left[e^{\Gamma(A_{1})}\right]$ as the partition function
of a system with Hamiltonian $-A_{1}$ at temperature $k_{B}T^{\circ}=1$.

We proceed to prove the formula for the product of two operators
\begin{equation}
{\rm Tr}\left[e^{\Gamma(A_{1})}e^{\Gamma(A_{2})}\right]={\rm det}\left(\mathbf{1}-\xi e^{A_{1}}e^{A_{2}}\right)^{-\xi}.\label{eq:TwoOperator}
\end{equation}
One can show that, the Fock space operators $\Gamma(A_{n})$ in Eq.
(\ref{eq:FockOperator}) satisfies

\begin{equation}
[\Gamma(A_{n}),\Gamma(A_{m})]=\Gamma([A_{n},A_{m}]),
\end{equation}
which implies for an $N$ dimensional single particle Hilbert space
$\Gamma$ is a representation of the usual Lie algebra of matrices
$gl(N)$. As a result, the Baker-Campbell-Hausdorf formula
\begin{equation}
e^{A_{1}}e^{A_{2}}=e^{B}
\end{equation}
leads to

\begin{equation}
e^{\Gamma(A_{1})}e^{\Gamma(A_{2})}=e^{\Gamma(B)}.
\end{equation}
Therefore, we have ${\rm Tr}\left[e^{\Gamma(A_{1})}e^{\Gamma(A_{2})}\right]={\rm Tr}[e^{\Gamma(B)}]={\rm det}\left(\mathbf{1}-\xi e^{B}\right)^{-\xi}={\rm det}\left(\mathbf{1}-\xi e^{A_{1}A_{2}}\right)^{-\xi}$
as shown in Eq. (\ref{eq:TwoOperator}). One can also see that this
relation can immediately be generalized in the same way to products
of more then two operators as our trace formula Eq. (\ref{eq:TraceFormula}).

A pedagogical example is the dimension of the Fock space whose coresponding
single particle Hilbert space has a dimension of $N$, which is given
by
\begin{equation}
{\rm Tr}\mathbf{1}={\rm Tr}e^{\Gamma(0)}={\rm det}(1-\xi)^{-\xi}=\begin{cases}
2^{N} & {\rm Fermions}\\
\infty & {\rm Bosons}
\end{cases},
\end{equation}
as it should be.

In this review, a commonly encounter case is that the last operator
$e^{\Gamma(A_{N})}$ in the trace formula is a fermion density matrix
in a bilinear form
\begin{equation}
\begin{aligned}\rho_{F} & =\frac{1}{Z}\exp\left(-\sum_{\alpha}\lambda_{\alpha}\hat{a}_{\alpha}^{\dagger}\hat{a}_{\alpha}\right),\\
 & \equiv\prod_{\alpha}\left[n_{\alpha}\hat{a}_{\alpha}^{\dagger}\hat{a}_{\alpha}+\left(1-n_{\alpha}\right)\hat{a}_{\alpha}\hat{a}_{\alpha}^{\dagger}\right],
\end{aligned}
\label{eq:rhoF}
\end{equation}
where 
\begin{equation}
e^{-\lambda_{\alpha}}=\frac{n_{\alpha}}{1-n_{\alpha}},
\end{equation}
with $n_{\alpha}$ being the distribution of fermions in the single
particle state corresponding to $\hat{a}_{\alpha}^{\dagger}$. The
normalized constant is given by $Z={\rm Tr}\exp\left(-\sum_{\alpha}\lambda_{\alpha}\hat{a}_{\alpha}^{\dagger}\hat{a}_{\alpha}\right)={\rm det}[\left(1-\hat{n}\right)^{-1}]$,
where $\hat{n}$ is a diagonal matrix with matrix elements $n_{\alpha}$.

One familiar example is the non-interacting Fermions at a the finite
temperature, where $\hat{a}_{\alpha}^{\dagger}$ creates a fermion
in the state with single-particle energy $\epsilon_{\alpha}$ and
\begin{equation}
n_{\alpha}=\frac{1}{e^{\left(\epsilon_{\alpha}-\mu\right)/k_{B}T^{\circ}}+1},\ \lambda_{\alpha}=\frac{\epsilon_{\alpha}-\mu}{k_{B}T^{\circ}}.
\end{equation}
Here, $\mu$ is the chemical potential, and $k_{B}$ is the Boltzmann
constant.

In this case, we have

\begin{equation}
\begin{aligned}{\rm Tr}\left[e^{\Gamma(A_{1})}e^{\Gamma(A_{2})}...e^{\Gamma(A_{N-1})}\rho_{F}\right]= & \frac{1}{Z}{\rm det}[\mathbf{1}+e^{A_{1}}e^{A_{2}}...\\
 & \times e^{A_{N-1}}e^{-\underline{\lambda}}]
\end{aligned}
\end{equation}
where in the basis of single particle states corresponding to $a_{\alpha}^{\dagger}$,
$e^{-\underline{\lambda}}$ is a diagonal matrix with matrix elements
$e^{-\lambda_{\alpha}}$, which leads to

\begin{equation}
e^{-\underline{\lambda}}=\hat{n}\left(\mathbf{1}-\hat{n}\right)^{-1}.
\end{equation}
Inserting the expression of $e^{-\underline{\lambda}}$ and $Z$ in
terms of distribution matrix $\hat{n}$ gives
\begin{equation}
{\rm Tr}\left[e^{\Gamma(A_{1})}e^{\Gamma(A_{2})}...e^{\Gamma(A_{N-1})}\rho_{F}\right]=\det[(\mathbf{1}-\hat{n})+\hat{R}\hat{n}],\label{eq:FDA}
\end{equation}
where

\begin{equation}
\hat{R}=e^{A_{1}}e^{A_{2}}...e^{A_{N-1}}.
\end{equation}

Another closely related and useful trace formula is
\begin{equation}
\begin{aligned}{\rm Tr}\left[e^{\Gamma(A_{1})}e^{\Gamma(A_{2})}...e^{\Gamma(A_{N})}a_{i}^{\dagger}a_{j}\right]= & (\mathbf{1}-\xi e^{-W})_{ji}^{-1}\\
 & \times{\rm det}\left(\mathbf{1}-\xi e^{W}\right)^{-\xi}
\end{aligned}
\label{eq:TraceFormula2}
\end{equation}
where $e^{W}=e^{A_{1}}e^{A_{2}}...e^{A_{N}}$. Noticing that ${\rm Tr}\left[e^{\Gamma(A_{1})}e^{\Gamma(A_{2})}...e^{\Gamma(A_{N})}\right]={\rm Tr}[e^{\Gamma(W)}]$
leads to
\begin{equation}
{\rm Tr}\left[e^{\Gamma(A_{1})}e^{\Gamma(A_{2})}...e^{\Gamma(A_{N})}a_{i}^{\dagger}a_{j}\right]=\frac{\partial}{\partial W_{ij}}{\rm Tr}[e^{\Gamma(W)}],
\end{equation}
where $W_{ij}\equiv\langle i|W|j\rangle$ are the matrix element of
$W$. Applying Jacobi's formula gives
\begin{equation}
\frac{\partial{\rm det}(\mathbf{1}-\xi e^{W})^{-\xi}}{\partial W_{ij}}={\rm {\rm det}}(\mathbf{1}-\xi e^{W})^{-\xi}{\rm Tr}[(\mathbf{1}-\xi e^{-W})^{-1}\frac{\partial W}{\partial W_{ij}}],
\end{equation}
where taking the trace on the right-hand-side gives ${\rm Tr}[(\mathbf{1}-\xi e^{-W})^{-1}\partial W/\partial W_{ij}]=(\mathbf{1}-\xi e^{-W})_{ji}^{-1}$,
which evntually leads to Eq. (\ref{eq:TraceFormula2}).

\section{Numerical Calculations}

Numerical calculations are carried out in a finite system confined
in a sphere of radius $R$. Keeping the density constant, we increase
$R$ towards infinity until numerical results are converged. Typically,
we choose $k_{F}R=250\pi$ in a calculation. We focus on the $s$-wave
interaction channel between $|\uparrow\rangle$ and the background
fermions near a broad Feshbach resonance, which can be well mimicked
by a spherically symmetric and short-range van-der-Waals type potential
$V(r)=-C_{6}\exp(-r_{0}^{6}/r^{6})/r^{6}$ \cite{Jia2012PRL,YujunPRL2012,JiaWangPRA2012}.
Here, $C_{6}$ determines the van-der-Waals length $l_{{\rm vdW}}=(2mC_{6})^{1/4}/2$,
and we choose $l_{{\rm vdW}}k_{F}=0.01\ll1$, so the short-range details
are unimportant. $r_{0}$ is the short-range parameter that tunes
the scattering length $a$. We choose $k_{F}r_{0}\approx7\times10^{-3}$,
which can support two bound states on the positive side. We also include
about $1100$ continuum states in a typical calculation. Covergence
with respect to both number of bound states and continnum states have
been tested.

\bibliography{FDA_Review}

\begin{thebibliography}{79}%
\makeatletter
\providecommand \@ifxundefined [1]{%
 \@ifx{#1\undefined}
}%
\providecommand \@ifnum [1]{%
 \ifnum #1\expandafter \@firstoftwo
 \else \expandafter \@secondoftwo
 \fi
}%
\providecommand \@ifx [1]{%
 \ifx #1\expandafter \@firstoftwo
 \else \expandafter \@secondoftwo
 \fi
}%
\providecommand \natexlab [1]{#1}%
\providecommand \enquote  [1]{``#1''}%
\providecommand \bibnamefont  [1]{#1}%
\providecommand \bibfnamefont [1]{#1}%
\providecommand \citenamefont [1]{#1}%
\providecommand \href@noop [0]{\@secondoftwo}%
\providecommand \href [0]{\begingroup \@sanitize@url \@href}%
\providecommand \@href[1]{\@@startlink{#1}\@@href}%
\providecommand \@@href[1]{\endgroup#1\@@endlink}%
\providecommand \@sanitize@url [0]{\catcode `\\12\catcode `\$12\catcode
  `\&12\catcode `\#12\catcode `\^12\catcode `\_12\catcode `\%12\relax}%
\providecommand \@@startlink[1]{}%
\providecommand \@@endlink[0]{}%
\providecommand \url  [0]{\begingroup\@sanitize@url \@url }%
\providecommand \@url [1]{\endgroup\@href {#1}{\urlprefix }}%
\providecommand \urlprefix  [0]{URL }%
\providecommand \Eprint [0]{\href }%
\providecommand \doibase [0]{http://dx.doi.org/}%
\providecommand \selectlanguage [0]{\@gobble}%
\providecommand \bibinfo  [0]{\@secondoftwo}%
\providecommand \bibfield  [0]{\@secondoftwo}%
\providecommand \translation [1]{[#1]}%
\providecommand \BibitemOpen [0]{}%
\providecommand \bibitemStop [0]{}%
\providecommand \bibitemNoStop [0]{.\EOS\space}%
\providecommand \EOS [0]{\spacefactor3000\relax}%
\providecommand \BibitemShut  [1]{\csname bibitem#1\endcsname}%
\let\auto@bib@innerbib\@empty
\bibitem [{\citenamefont {Mahan}(2000)}]{Mahan2000Book}%
  \BibitemOpen
  \bibfield  {author} {\bibinfo {author} {\bibfnamefont {Gerald~D.}\
  \bibnamefont {Mahan}},\ }\href@noop {} {\emph {\bibinfo {title} {Many
  Particle Physics}}},\ \bibinfo {edition} {3rd}\ ed.\ (\bibinfo  {publisher}
  {Kluwer},\ \bibinfo {address} {New York},\ \bibinfo {year}
  {2000})\BibitemShut {NoStop}%
\bibitem [{\citenamefont {Mahan}(1967{\natexlab{a}})}]{Mahan1967PR1}%
  \BibitemOpen
  \bibfield  {author} {\bibinfo {author} {\bibfnamefont {G.~D.}\ \bibnamefont
  {Mahan}},\ }\bibfield  {title} {\enquote {\bibinfo {title} {Excitons in
  degenerate semiconductors},}\ }\href@noop {} {\bibfield  {journal} {\bibinfo
  {journal} {Phys. Rev.}\ }\textbf {\bibinfo {volume} {153}},\ \bibinfo {pages}
  {882--889} (\bibinfo {year} {1967}{\natexlab{a}})}\BibitemShut {NoStop}%
\bibitem [{\citenamefont {Mahan}(1967{\natexlab{b}})}]{Mahan1967PR2}%
  \BibitemOpen
  \bibfield  {author} {\bibinfo {author} {\bibfnamefont {G.~D.}\ \bibnamefont
  {Mahan}},\ }\bibfield  {title} {\enquote {\bibinfo {title} {Excitons in
  metals: Infinite hole mass},}\ }\href@noop {} {\bibfield  {journal} {\bibinfo
   {journal} {Phys. Rev.}\ }\textbf {\bibinfo {volume} {163}},\ \bibinfo
  {pages} {612--617} (\bibinfo {year} {1967}{\natexlab{b}})}\BibitemShut
  {NoStop}%
\bibitem [{\citenamefont {Nozi{\`e}res}\ and\ \citenamefont {{De
  Dominics}}(1969)}]{Nozieres1969PR}%
  \BibitemOpen
  \bibfield  {author} {\bibinfo {author} {\bibfnamefont {P.}~\bibnamefont
  {Nozi{\`e}res}}\ and\ \bibinfo {author} {\bibfnamefont {C.~T.}\ \bibnamefont
  {{De Dominics}}},\ }\bibfield  {title} {\enquote {\bibinfo {title}
  {Singularities in the x-ray absorption and emission of metals. iii. one-body
  theory exact solution},}\ }\href@noop {} {\bibfield  {journal} {\bibinfo
  {journal} {Phys. Rev.}\ }\textbf {\bibinfo {volume} {178}},\ \bibinfo {pages}
  {1097--1107} (\bibinfo {year} {1969})}\BibitemShut {NoStop}%
\bibitem [{\citenamefont {Anderson}(1967)}]{Anderson1967PRL}%
  \BibitemOpen
  \bibfield  {author} {\bibinfo {author} {\bibfnamefont {P.~W.}\ \bibnamefont
  {Anderson}},\ }\bibfield  {title} {\enquote {\bibinfo {title} {Infrared
  catastrophe in fermi gases with local scattering potentials},}\ }\href@noop
  {} {\bibfield  {journal} {\bibinfo  {journal} {Phys. Rev. Lett.}\ }\textbf
  {\bibinfo {volume} {18}},\ \bibinfo {pages} {1049--1051} (\bibinfo {year}
  {1967})}\BibitemShut {NoStop}%
\bibitem [{\citenamefont {Matveev}\ and\ \citenamefont
  {Larkin}(1992)}]{Matveev1992PRB}%
  \BibitemOpen
  \bibfield  {author} {\bibinfo {author} {\bibfnamefont {K.~A.}\ \bibnamefont
  {Matveev}}\ and\ \bibinfo {author} {\bibfnamefont {A.~I.}\ \bibnamefont
  {Larkin}},\ }\bibfield  {title} {\enquote {\bibinfo {title}
  {Interaction-induced threshold singularities in tunneling via localized
  levels},}\ }\href@noop {} {\bibfield  {journal} {\bibinfo  {journal} {Phys.
  Rev. B}\ }\textbf {\bibinfo {volume} {46}},\ \bibinfo {pages} {15337--15347}
  (\bibinfo {year} {1992})}\BibitemShut {NoStop}%
\bibitem [{\citenamefont {Geim}\ \emph {et~al.}(1994)\citenamefont {Geim},
  \citenamefont {Main}, \citenamefont {La~Scala}, \citenamefont {Eaves},
  \citenamefont {Foster}, \citenamefont {Beton}, \citenamefont {Sakai},
  \citenamefont {Sheard}, \citenamefont {Henini}, \citenamefont {Hill},\ and\
  \citenamefont {Pate}}]{Pate1994PRL}%
  \BibitemOpen
  \bibfield  {author} {\bibinfo {author} {\bibfnamefont {A.~K.}\ \bibnamefont
  {Geim}}, \bibinfo {author} {\bibfnamefont {P.~C.}\ \bibnamefont {Main}},
  \bibinfo {author} {\bibfnamefont {N.}~\bibnamefont {La~Scala}}, \bibinfo
  {author} {\bibfnamefont {L.}~\bibnamefont {Eaves}}, \bibinfo {author}
  {\bibfnamefont {T.~J.}\ \bibnamefont {Foster}}, \bibinfo {author}
  {\bibfnamefont {P.~H.}\ \bibnamefont {Beton}}, \bibinfo {author}
  {\bibfnamefont {J.~W.}\ \bibnamefont {Sakai}}, \bibinfo {author}
  {\bibfnamefont {F.~W.}\ \bibnamefont {Sheard}}, \bibinfo {author}
  {\bibfnamefont {M.}~\bibnamefont {Henini}}, \bibinfo {author} {\bibfnamefont
  {G.}~\bibnamefont {Hill}}, \ and\ \bibinfo {author} {\bibfnamefont {M.~A.}\
  \bibnamefont {Pate}},\ }\bibfield  {title} {\enquote {\bibinfo {title}
  {Fermi-edge singularity in resonant tunneling},}\ }\href@noop {} {\bibfield
  {journal} {\bibinfo  {journal} {Phys. Rev. Lett.}\ }\textbf {\bibinfo
  {volume} {72}},\ \bibinfo {pages} {2061--2064} (\bibinfo {year}
  {1994})}\BibitemShut {NoStop}%
\bibitem [{\citenamefont {Ogawa}\ \emph {et~al.}(1992)\citenamefont {Ogawa},
  \citenamefont {Furusaki},\ and\ \citenamefont {Nagaosa}}]{Nagaosa1992PRL}%
  \BibitemOpen
  \bibfield  {author} {\bibinfo {author} {\bibfnamefont {Tetsuo}\ \bibnamefont
  {Ogawa}}, \bibinfo {author} {\bibfnamefont {Akira}\ \bibnamefont {Furusaki}},
  \ and\ \bibinfo {author} {\bibfnamefont {Naoto}\ \bibnamefont {Nagaosa}},\
  }\bibfield  {title} {\enquote {\bibinfo {title} {Fermi-edge singularity in
  one-dimensional systems},}\ }\href@noop {} {\bibfield  {journal} {\bibinfo
  {journal} {Phys. Rev. Lett.}\ }\textbf {\bibinfo {volume} {68}},\ \bibinfo
  {pages} {3638--3641} (\bibinfo {year} {1992})}\BibitemShut {NoStop}%
\bibitem [{\citenamefont {Prokof'ev}(1994)}]{Prokof'ev1994PRB}%
  \BibitemOpen
  \bibfield  {author} {\bibinfo {author} {\bibfnamefont {N.~V.}\ \bibnamefont
  {Prokof'ev}},\ }\bibfield  {title} {\enquote {\bibinfo {title} {Fermi-edge
  singularity with backscattering in the luttinger-liquid model},}\ }\href@noop
  {} {\bibfield  {journal} {\bibinfo  {journal} {Phys. Rev. B}\ }\textbf
  {\bibinfo {volume} {49}},\ \bibinfo {pages} {2148--2151} (\bibinfo {year}
  {1994})}\BibitemShut {NoStop}%
\bibitem [{\citenamefont {Komnik}\ \emph {et~al.}(1997)\citenamefont {Komnik},
  \citenamefont {Egger},\ and\ \citenamefont {Gogolin}}]{Komnik1997PRB}%
  \BibitemOpen
  \bibfield  {author} {\bibinfo {author} {\bibfnamefont {Andrei}\ \bibnamefont
  {Komnik}}, \bibinfo {author} {\bibfnamefont {Reinhold}\ \bibnamefont
  {Egger}}, \ and\ \bibinfo {author} {\bibfnamefont {Alexander~O.}\
  \bibnamefont {Gogolin}},\ }\bibfield  {title} {\enquote {\bibinfo {title}
  {Exact fermi-edge singularity exponent in a luttinger liquid},}\ }\href@noop
  {} {\bibfield  {journal} {\bibinfo  {journal} {Phys. Rev. B}\ }\textbf
  {\bibinfo {volume} {56}},\ \bibinfo {pages} {1153--1160} (\bibinfo {year}
  {1997})}\BibitemShut {NoStop}%
\bibitem [{\citenamefont {Bascones}\ \emph {et~al.}(2000)\citenamefont
  {Bascones}, \citenamefont {Herrero}, \citenamefont {Guinea},\ and\
  \citenamefont {Sch\"on}}]{bascones2000PRB}%
  \BibitemOpen
  \bibfield  {author} {\bibinfo {author} {\bibfnamefont {E.}~\bibnamefont
  {Bascones}}, \bibinfo {author} {\bibfnamefont {C.~P.}\ \bibnamefont
  {Herrero}}, \bibinfo {author} {\bibfnamefont {F.}~\bibnamefont {Guinea}}, \
  and\ \bibinfo {author} {\bibfnamefont {Gerd}\ \bibnamefont {Sch\"on}},\
  }\bibfield  {title} {\enquote {\bibinfo {title} {Nonequilibrium effects in
  transport through quantum dots},}\ }\href@noop {} {\bibfield  {journal}
  {\bibinfo  {journal} {Phys. Rev. B}\ }\textbf {\bibinfo {volume} {61}},\
  \bibinfo {pages} {16778--16786} (\bibinfo {year} {2000})}\BibitemShut
  {NoStop}%
\bibitem [{\citenamefont {Levitov}\ and\ \citenamefont
  {Lee}(1996)}]{Leonid1996JMathPhys}%
  \BibitemOpen
  \bibfield  {author} {\bibinfo {author} {\bibfnamefont {Leonid~S.}\
  \bibnamefont {Levitov}}\ and\ \bibinfo {author} {\bibfnamefont {Hyunwoo}\
  \bibnamefont {Lee}},\ }\bibfield  {title} {\enquote {\bibinfo {title}
  {Electron counting statistics and coherent states of electric current},}\
  }\href@noop {} {\bibfield  {journal} {\bibinfo  {journal} {J. Math. Phys.}\
  }\textbf {\bibinfo {volume} {37}},\ \bibinfo {pages} {4845} (\bibinfo {year}
  {1996})}\BibitemShut {NoStop}%
\bibitem [{\citenamefont {Klich}(2003)}]{Klich2003Book}%
  \BibitemOpen
  \bibfield  {author} {\bibinfo {author} {\bibfnamefont {I.}~\bibnamefont
  {Klich}},\ }\href@noop {} {\emph {\bibinfo {title} {Full Counting Statistics:
  an Elementary Derivation of {Levitov's} Formula}}}\ (\bibinfo  {publisher}
  {Kluwer},\ \bibinfo {address} {Dordrecht},\ \bibinfo {year}
  {2003})\BibitemShut {NoStop}%
\bibitem [{\citenamefont {Sch{\"o}nhammer}(2007)}]{Schonhammer2007PRB}%
  \BibitemOpen
  \bibfield  {author} {\bibinfo {author} {\bibfnamefont {K.}~\bibnamefont
  {Sch{\"o}nhammer}},\ }\bibfield  {title} {\enquote {\bibinfo {title} {Full
  counting statistics for noninteracting fermions: Exact results and the
  {Levitov-Lesovik} formula},}\ }\href@noop {} {\bibfield  {journal} {\bibinfo
  {journal} {Phys. Rev. B}\ }\textbf {\bibinfo {volume} {75}},\ \bibinfo
  {pages} {205329} (\bibinfo {year} {2007})}\BibitemShut {NoStop}%
\bibitem [{\citenamefont {Ivanov}\ and\ \citenamefont
  {Abanov}(2013)}]{Ivanov2013JMathPhys}%
  \BibitemOpen
  \bibfield  {author} {\bibinfo {author} {\bibfnamefont {Dmitri~A}\
  \bibnamefont {Ivanov}}\ and\ \bibinfo {author} {\bibfnamefont {Alexander~G}\
  \bibnamefont {Abanov}},\ }\bibfield  {title} {\enquote {\bibinfo {title}
  {{Fisher-Hartwig} expansion for {Toeplitz} determinants and the spectrum of a
  single-particle reduced density matrix for one-dimensional free fermions},}\
  }\href@noop {} {\bibfield  {journal} {\bibinfo  {journal} {J. Phys. A: Math.
  Theor.}\ }\textbf {\bibinfo {volume} {46}},\ \bibinfo {pages} {375005}
  (\bibinfo {year} {2013})}\BibitemShut {NoStop}%
\bibitem [{\citenamefont {Muzykantskii}\ \emph {et~al.}(2003)\citenamefont
  {Muzykantskii}, \citenamefont {d'Ambrumenil},\ and\ \citenamefont
  {Braunecker}}]{Muzykantskii2003PRL}%
  \BibitemOpen
  \bibfield  {author} {\bibinfo {author} {\bibfnamefont {B.}~\bibnamefont
  {Muzykantskii}}, \bibinfo {author} {\bibfnamefont {N.}~\bibnamefont
  {d'Ambrumenil}}, \ and\ \bibinfo {author} {\bibfnamefont {B.}~\bibnamefont
  {Braunecker}},\ }\bibfield  {title} {\enquote {\bibinfo {title} {Fermi-edge
  singularity in a nonequilibrium system},}\ }\href@noop {} {\bibfield
  {journal} {\bibinfo  {journal} {Phys. Rev. Lett.}\ }\textbf {\bibinfo
  {volume} {91}},\ \bibinfo {pages} {266602} (\bibinfo {year}
  {2003})}\BibitemShut {NoStop}%
\bibitem [{\citenamefont {d'Ambrumenil}\ and\ \citenamefont
  {Muzykantskii}(2005)}]{Muzykantskii2005PRB}%
  \BibitemOpen
  \bibfield  {author} {\bibinfo {author} {\bibfnamefont {N.}~\bibnamefont
  {d'Ambrumenil}}\ and\ \bibinfo {author} {\bibfnamefont {B.}~\bibnamefont
  {Muzykantskii}},\ }\bibfield  {title} {\enquote {\bibinfo {title} {Fermi gas
  response to time-dependent perturbations},}\ }\href@noop {} {\bibfield
  {journal} {\bibinfo  {journal} {Phys. Rev. B}\ }\textbf {\bibinfo {volume}
  {71}},\ \bibinfo {pages} {045326} (\bibinfo {year} {2005})}\BibitemShut
  {NoStop}%
\bibitem [{\citenamefont {Abanin}\ and\ \citenamefont
  {Levitov}(2005)}]{Levitov2005PRL}%
  \BibitemOpen
  \bibfield  {author} {\bibinfo {author} {\bibfnamefont {D.~A.}\ \bibnamefont
  {Abanin}}\ and\ \bibinfo {author} {\bibfnamefont {L.~S.}\ \bibnamefont
  {Levitov}},\ }\bibfield  {title} {\enquote {\bibinfo {title} {Fermi-edge
  resonance and tunneling in nonequilibrium electron gas},}\ }\href@noop {}
  {\bibfield  {journal} {\bibinfo  {journal} {Phys. Rev. Lett.}\ }\textbf
  {\bibinfo {volume} {94}},\ \bibinfo {pages} {186803} (\bibinfo {year}
  {2005})}\BibitemShut {NoStop}%
\bibitem [{\citenamefont {Abanin}\ and\ \citenamefont
  {Levitov}(2004)}]{Levitov2004PRL}%
  \BibitemOpen
  \bibfield  {author} {\bibinfo {author} {\bibfnamefont {D.~A.}\ \bibnamefont
  {Abanin}}\ and\ \bibinfo {author} {\bibfnamefont {L.~S.}\ \bibnamefont
  {Levitov}},\ }\bibfield  {title} {\enquote {\bibinfo {title} {Tunable
  fermi-edge resonance in an open quantum dot},}\ }\href@noop {} {\bibfield
  {journal} {\bibinfo  {journal} {Phys. Rev. Lett.}\ }\textbf {\bibinfo
  {volume} {93}},\ \bibinfo {pages} {126802} (\bibinfo {year}
  {2004})}\BibitemShut {NoStop}%
\bibitem [{\citenamefont {Chang}\ and\ \citenamefont
  {Reichman}(2019)}]{Reichman2019PRB}%
  \BibitemOpen
  \bibfield  {author} {\bibinfo {author} {\bibfnamefont {Yao-Wen}\ \bibnamefont
  {Chang}}\ and\ \bibinfo {author} {\bibfnamefont {David~R.}\ \bibnamefont
  {Reichman}},\ }\bibfield  {title} {\enquote {\bibinfo {title} {Many-body
  theory of optical absorption in doped two-dimensional semiconductors},}\
  }\href@noop {} {\bibfield  {journal} {\bibinfo  {journal} {Phys. Rev. B}\
  }\textbf {\bibinfo {volume} {99}},\ \bibinfo {pages} {125421} (\bibinfo
  {year} {2019})}\BibitemShut {NoStop}%
\bibitem [{\citenamefont {Lindoy}\ \emph {et~al.}()\citenamefont {Lindoy},
  \citenamefont {Chang},\ and\ \citenamefont {Reichman}}]{Reichman2022arXiv}%
  \BibitemOpen
  \bibfield  {author} {\bibinfo {author} {\bibfnamefont {Lachlan~P}\
  \bibnamefont {Lindoy}}, \bibinfo {author} {\bibfnamefont {Yao-Wen}\
  \bibnamefont {Chang}}, \ and\ \bibinfo {author} {\bibfnamefont {David~R}\
  \bibnamefont {Reichman}},\ }\href@noop {} {\enquote {\bibinfo {title}
  {Two-dimensional spectroscopy of two-dimensional materials},}\ }\bibinfo
  {note} {{arXiv}:2206.01799 (2022)}\BibitemShut {NoStop}%
\bibitem [{\citenamefont {Knap}\ \emph {et~al.}(2012)\citenamefont {Knap},
  \citenamefont {Shashi}, \citenamefont {Nishida}, \citenamefont {Imambekov},
  \citenamefont {Abanin},\ and\ \citenamefont {Demler}}]{Demler2012PRX}%
  \BibitemOpen
  \bibfield  {author} {\bibinfo {author} {\bibfnamefont {Michael}\ \bibnamefont
  {Knap}}, \bibinfo {author} {\bibfnamefont {Aditya}\ \bibnamefont {Shashi}},
  \bibinfo {author} {\bibfnamefont {Yusuke}\ \bibnamefont {Nishida}}, \bibinfo
  {author} {\bibfnamefont {Adilet}\ \bibnamefont {Imambekov}}, \bibinfo
  {author} {\bibfnamefont {Dmitry~A.}\ \bibnamefont {Abanin}}, \ and\ \bibinfo
  {author} {\bibfnamefont {Eugene}\ \bibnamefont {Demler}},\ }\bibfield
  {title} {\enquote {\bibinfo {title} {Time-dependent impurity in ultracold
  fermions: Orthogonality catastrophe and beyond},}\ }\href@noop {} {\bibfield
  {journal} {\bibinfo  {journal} {Phys. Rev. X}\ }\textbf {\bibinfo {volume}
  {2}},\ \bibinfo {pages} {041020} (\bibinfo {year} {2012})}\BibitemShut
  {NoStop}%
\bibitem [{\citenamefont {Schmidt}\ \emph
  {et~al.}(2018{\natexlab{a}})\citenamefont {Schmidt}, \citenamefont {Knap},
  \citenamefont {Ivanov}, \citenamefont {You}, \citenamefont {Cetina},\ and\
  \citenamefont {Demler}}]{Schmidt2018Review}%
  \BibitemOpen
  \bibfield  {author} {\bibinfo {author} {\bibfnamefont {R.}~\bibnamefont
  {Schmidt}}, \bibinfo {author} {\bibfnamefont {M.}~\bibnamefont {Knap}},
  \bibinfo {author} {\bibfnamefont {D.~A.}\ \bibnamefont {Ivanov}}, \bibinfo
  {author} {\bibfnamefont {J.-S.}\ \bibnamefont {You}}, \bibinfo {author}
  {\bibfnamefont {M.}~\bibnamefont {Cetina}}, \ and\ \bibinfo {author}
  {\bibfnamefont {E.}~\bibnamefont {Demler}},\ }\bibfield  {title} {\enquote
  {\bibinfo {title} {Universal many-body response of heavy impurities coupled
  to a {Fermi} sea: a review of recent progress},}\ }\href@noop {} {\bibfield
  {journal} {\bibinfo  {journal} {Rep. Prog. Phys.}\ }\textbf {\bibinfo
  {volume} {81}},\ \bibinfo {pages} {024401} (\bibinfo {year}
  {2018}{\natexlab{a}})}\BibitemShut {NoStop}%
\bibitem [{\citenamefont {Goold}\ \emph {et~al.}(2011)\citenamefont {Goold},
  \citenamefont {Fogarty}, \citenamefont {Lo~Gullo}, \citenamefont
  {Paternostro},\ and\ \citenamefont {Busch}}]{Goold2011PRA}%
  \BibitemOpen
  \bibfield  {author} {\bibinfo {author} {\bibfnamefont {J.}~\bibnamefont
  {Goold}}, \bibinfo {author} {\bibfnamefont {T.}~\bibnamefont {Fogarty}},
  \bibinfo {author} {\bibfnamefont {N.}~\bibnamefont {Lo~Gullo}}, \bibinfo
  {author} {\bibfnamefont {M.}~\bibnamefont {Paternostro}}, \ and\ \bibinfo
  {author} {\bibfnamefont {Th.}\ \bibnamefont {Busch}},\ }\bibfield  {title}
  {\enquote {\bibinfo {title} {Orthogonality catastrophe as a consequence of
  qubit embedding in an ultracold fermi gas},}\ }\href@noop {} {\bibfield
  {journal} {\bibinfo  {journal} {Phys. Rev. A}\ }\textbf {\bibinfo {volume}
  {84}},\ \bibinfo {pages} {063632} (\bibinfo {year} {2011})}\BibitemShut
  {NoStop}%
\bibitem [{\citenamefont {Chevy}(2006)}]{Chevy2006PRA}%
  \BibitemOpen
  \bibfield  {author} {\bibinfo {author} {\bibfnamefont {F.}~\bibnamefont
  {Chevy}},\ }\bibfield  {title} {\enquote {\bibinfo {title} {Universal phase
  diagram of a strongly interacting {Fermi} gas with unbalanced spin
  populations},}\ }\href@noop {} {\bibfield  {journal} {\bibinfo  {journal}
  {Phys. Rev. A}\ }\textbf {\bibinfo {volume} {74}},\ \bibinfo {pages} {063628}
  (\bibinfo {year} {2006})}\BibitemShut {NoStop}%
\bibitem [{\citenamefont {Combescot}\ \emph {et~al.}(2007)\citenamefont
  {Combescot}, \citenamefont {Recati}, \citenamefont {Lobo},\ and\
  \citenamefont {Chevy}}]{Combescot2007PRL}%
  \BibitemOpen
  \bibfield  {author} {\bibinfo {author} {\bibfnamefont {R.}~\bibnamefont
  {Combescot}}, \bibinfo {author} {\bibfnamefont {A.}~\bibnamefont {Recati}},
  \bibinfo {author} {\bibfnamefont {C.}~\bibnamefont {Lobo}}, \ and\ \bibinfo
  {author} {\bibfnamefont {F.}~\bibnamefont {Chevy}},\ }\bibfield  {title}
  {\enquote {\bibinfo {title} {Normal state of highly polarized {Fermi} gases:
  Simple many-body approaches},}\ }\href@noop {} {\bibfield  {journal}
  {\bibinfo  {journal} {Phys. Rev. Lett.}\ }\textbf {\bibinfo {volume} {98}},\
  \bibinfo {pages} {180402} (\bibinfo {year} {2007})}\BibitemShut {NoStop}%
\bibitem [{\citenamefont {Punk}\ \emph {et~al.}(2009)\citenamefont {Punk},
  \citenamefont {Dumitrescu},\ and\ \citenamefont {Zwerger}}]{Punk2009PRA}%
  \BibitemOpen
  \bibfield  {author} {\bibinfo {author} {\bibfnamefont {M.}~\bibnamefont
  {Punk}}, \bibinfo {author} {\bibfnamefont {P.~T.}\ \bibnamefont
  {Dumitrescu}}, \ and\ \bibinfo {author} {\bibfnamefont {W.}~\bibnamefont
  {Zwerger}},\ }\bibfield  {title} {\enquote {\bibinfo {title}
  {Polaron-to-molecule transition in a strongly imbalanced {Fermi} gas},}\
  }\href@noop {} {\bibfield  {journal} {\bibinfo  {journal} {Phys. Rev. A}\
  }\textbf {\bibinfo {volume} {80}},\ \bibinfo {pages} {053605} (\bibinfo
  {year} {2009})}\BibitemShut {NoStop}%
\bibitem [{\citenamefont {Cui}\ and\ \citenamefont {Zhai}(2010)}]{Cui2010PRA}%
  \BibitemOpen
  \bibfield  {author} {\bibinfo {author} {\bibfnamefont {Xiaoling}\
  \bibnamefont {Cui}}\ and\ \bibinfo {author} {\bibfnamefont {Hui}\
  \bibnamefont {Zhai}},\ }\bibfield  {title} {\enquote {\bibinfo {title}
  {Stability of a fully magnetized ferromagnetic state in repulsively
  interacting ultracold {Fermi} gases},}\ }\href@noop {} {\bibfield  {journal}
  {\bibinfo  {journal} {Phys. Rev. A}\ }\textbf {\bibinfo {volume} {81}},\
  \bibinfo {pages} {041602} (\bibinfo {year} {2010})}\BibitemShut {NoStop}%
\bibitem [{\citenamefont {Mathy}\ \emph {et~al.}(2011)\citenamefont {Mathy},
  \citenamefont {Parish},\ and\ \citenamefont {Huse}}]{Mathy2011PRL}%
  \BibitemOpen
  \bibfield  {author} {\bibinfo {author} {\bibfnamefont {Charles J.~M.}\
  \bibnamefont {Mathy}}, \bibinfo {author} {\bibfnamefont {Meera~M.}\
  \bibnamefont {Parish}}, \ and\ \bibinfo {author} {\bibfnamefont {David~A.}\
  \bibnamefont {Huse}},\ }\bibfield  {title} {\enquote {\bibinfo {title}
  {Trimers, molecules, and polarons in mass-imbalanced atomic {Fermi} gases},}\
  }\href@noop {} {\bibfield  {journal} {\bibinfo  {journal} {Phys. Rev. Lett.}\
  }\textbf {\bibinfo {volume} {106}},\ \bibinfo {pages} {166404} (\bibinfo
  {year} {2011})}\BibitemShut {NoStop}%
\bibitem [{\citenamefont {Schmidt}\ \emph {et~al.}(2012)\citenamefont
  {Schmidt}, \citenamefont {Enss}, \citenamefont {Pietil\"a},\ and\
  \citenamefont {Demler}}]{Schmidt2012PRA}%
  \BibitemOpen
  \bibfield  {author} {\bibinfo {author} {\bibfnamefont {Richard}\ \bibnamefont
  {Schmidt}}, \bibinfo {author} {\bibfnamefont {Tilman}\ \bibnamefont {Enss}},
  \bibinfo {author} {\bibfnamefont {Ville}\ \bibnamefont {Pietil\"a}}, \ and\
  \bibinfo {author} {\bibfnamefont {Eugene}\ \bibnamefont {Demler}},\
  }\bibfield  {title} {\enquote {\bibinfo {title} {Fermi polarons in two
  dimensions},}\ }\href@noop {} {\bibfield  {journal} {\bibinfo  {journal}
  {Phys. Rev. A}\ }\textbf {\bibinfo {volume} {85}},\ \bibinfo {pages} {021602}
  (\bibinfo {year} {2012})}\BibitemShut {NoStop}%
\bibitem [{\citenamefont {Parish}\ and\ \citenamefont
  {Levinsen}(2013)}]{Parish2013PRA}%
  \BibitemOpen
  \bibfield  {author} {\bibinfo {author} {\bibfnamefont {Meera~M.}\
  \bibnamefont {Parish}}\ and\ \bibinfo {author} {\bibfnamefont {Jesper}\
  \bibnamefont {Levinsen}},\ }\bibfield  {title} {\enquote {\bibinfo {title}
  {Highly polarized fermi gases in two dimensions},}\ }\href@noop {} {\bibfield
   {journal} {\bibinfo  {journal} {Phys. Rev. A}\ }\textbf {\bibinfo {volume}
  {87}},\ \bibinfo {pages} {033616} (\bibinfo {year} {2013})}\BibitemShut
  {NoStop}%
\bibitem [{\citenamefont {Levinsen}\ \emph {et~al.}(2015)\citenamefont
  {Levinsen}, \citenamefont {Parish},\ and\ \citenamefont
  {Bruun}}]{Levinsen2015PRL}%
  \BibitemOpen
  \bibfield  {author} {\bibinfo {author} {\bibfnamefont {Jesper}\ \bibnamefont
  {Levinsen}}, \bibinfo {author} {\bibfnamefont {Meera~M.}\ \bibnamefont
  {Parish}}, \ and\ \bibinfo {author} {\bibfnamefont {Georg~M.}\ \bibnamefont
  {Bruun}},\ }\bibfield  {title} {\enquote {\bibinfo {title} {Impurity in a
  bose-einstein condensate and the efimov effect},}\ }\href@noop {} {\bibfield
  {journal} {\bibinfo  {journal} {Phys. Rev. Lett.}\ }\textbf {\bibinfo
  {volume} {115}},\ \bibinfo {pages} {125302} (\bibinfo {year}
  {2015})}\BibitemShut {NoStop}%
\bibitem [{\citenamefont {Hu}\ \emph {et~al.}(2016)\citenamefont {Hu},
  \citenamefont {Wang}, \citenamefont {Yi},\ and\ \citenamefont
  {Liu}}]{HuHui2016PRA}%
  \BibitemOpen
  \bibfield  {author} {\bibinfo {author} {\bibfnamefont {Hui}\ \bibnamefont
  {Hu}}, \bibinfo {author} {\bibfnamefont {An-Bang}\ \bibnamefont {Wang}},
  \bibinfo {author} {\bibfnamefont {Su}~\bibnamefont {Yi}}, \ and\ \bibinfo
  {author} {\bibfnamefont {Xia-Ji}\ \bibnamefont {Liu}},\ }\bibfield  {title}
  {\enquote {\bibinfo {title} {Fermi polaron in a one-dimensional quasiperiodic
  optical lattice: The simplest many-body localization challenge},}\
  }\href@noop {} {\bibfield  {journal} {\bibinfo  {journal} {Phys. Rev. A}\
  }\textbf {\bibinfo {volume} {93}},\ \bibinfo {pages} {053601} (\bibinfo
  {year} {2016})}\BibitemShut {NoStop}%
\bibitem [{\citenamefont {Hu}\ \emph {et~al.}(2018)\citenamefont {Hu},
  \citenamefont {Mulkerin}, \citenamefont {Wang},\ and\ \citenamefont
  {Liu}}]{HuHui2018PRA}%
  \BibitemOpen
  \bibfield  {author} {\bibinfo {author} {\bibfnamefont {Hui}\ \bibnamefont
  {Hu}}, \bibinfo {author} {\bibfnamefont {Brendan~C.}\ \bibnamefont
  {Mulkerin}}, \bibinfo {author} {\bibfnamefont {Jia}\ \bibnamefont {Wang}}, \
  and\ \bibinfo {author} {\bibfnamefont {Xia-Ji}\ \bibnamefont {Liu}},\
  }\bibfield  {title} {\enquote {\bibinfo {title} {Attractive fermi polarons at
  nonzero temperatures with a finite impurity concentration},}\ }\href@noop {}
  {\bibfield  {journal} {\bibinfo  {journal} {Phys. Rev. A}\ }\textbf {\bibinfo
  {volume} {98}},\ \bibinfo {pages} {013626} (\bibinfo {year}
  {2018})}\BibitemShut {NoStop}%
\bibitem [{\citenamefont {Mulkerin}\ \emph {et~al.}(2019)\citenamefont
  {Mulkerin}, \citenamefont {Liu},\ and\ \citenamefont
  {Hu}}]{Mulkerin2019AnnPhys}%
  \BibitemOpen
  \bibfield  {author} {\bibinfo {author} {\bibfnamefont {B.~C.}\ \bibnamefont
  {Mulkerin}}, \bibinfo {author} {\bibfnamefont {X.-J.}\ \bibnamefont {Liu}}, \
  and\ \bibinfo {author} {\bibfnamefont {H.}~\bibnamefont {Hu}},\ }\bibfield
  {title} {\enquote {\bibinfo {title} {Breakdown of the fermi polaron
  description near fermi degeneracy at unitarity},}\ }\href@noop {} {\bibfield
  {journal} {\bibinfo  {journal} {Ann. Phys. (NY)}\ }\textbf {\bibinfo {volume}
  {407}},\ \bibinfo {pages} {29} (\bibinfo {year} {2019})}\BibitemShut
  {NoStop}%
\bibitem [{\citenamefont {Parish}\ \emph {et~al.}(2021)\citenamefont {Parish},
  \citenamefont {Adlong}, \citenamefont {Liu},\ and\ \citenamefont
  {Levinsen}}]{Parish2021PRA}%
  \BibitemOpen
  \bibfield  {author} {\bibinfo {author} {\bibfnamefont {Meera~M.}\
  \bibnamefont {Parish}}, \bibinfo {author} {\bibfnamefont {Haydn~S.}\
  \bibnamefont {Adlong}}, \bibinfo {author} {\bibfnamefont {Weizhe~Edward}\
  \bibnamefont {Liu}}, \ and\ \bibinfo {author} {\bibfnamefont {Jesper}\
  \bibnamefont {Levinsen}},\ }\bibfield  {title} {\enquote {\bibinfo {title}
  {Thermodynamic signatures of the polaron-molecule transition in a fermi
  gas},}\ }\href@noop {} {\bibfield  {journal} {\bibinfo  {journal} {Phys. Rev.
  A}\ }\textbf {\bibinfo {volume} {103}},\ \bibinfo {pages} {023312} (\bibinfo
  {year} {2021})}\BibitemShut {NoStop}%
\bibitem [{\citenamefont {Lobo}\ \emph {et~al.}(2006)\citenamefont {Lobo},
  \citenamefont {Recati}, \citenamefont {Giorgini},\ and\ \citenamefont
  {Stringari}}]{Lobo2006PRL}%
  \BibitemOpen
  \bibfield  {author} {\bibinfo {author} {\bibfnamefont {C.}~\bibnamefont
  {Lobo}}, \bibinfo {author} {\bibfnamefont {A.}~\bibnamefont {Recati}},
  \bibinfo {author} {\bibfnamefont {S.}~\bibnamefont {Giorgini}}, \ and\
  \bibinfo {author} {\bibfnamefont {S.}~\bibnamefont {Stringari}},\ }\bibfield
  {title} {\enquote {\bibinfo {title} {Normal state of a polarized {Fermi} gas
  at unitarity},}\ }\href@noop {} {\bibfield  {journal} {\bibinfo  {journal}
  {Phys. Rev. Lett.}\ }\textbf {\bibinfo {volume} {97}},\ \bibinfo {pages}
  {200403} (\bibinfo {year} {2006})}\BibitemShut {NoStop}%
\bibitem [{\citenamefont {Kroiss}\ and\ \citenamefont
  {Pollet}(2015)}]{Kroiss2015PRL}%
  \BibitemOpen
  \bibfield  {author} {\bibinfo {author} {\bibfnamefont {Peter}\ \bibnamefont
  {Kroiss}}\ and\ \bibinfo {author} {\bibfnamefont {Lode}\ \bibnamefont
  {Pollet}},\ }\bibfield  {title} {\enquote {\bibinfo {title} {Diagrammatic
  monte carlo study of a mass-imbalanced fermi-polaron system},}\ }\href@noop
  {} {\bibfield  {journal} {\bibinfo  {journal} {Phys. Rev. B}\ }\textbf
  {\bibinfo {volume} {91}},\ \bibinfo {pages} {144507} (\bibinfo {year}
  {2015})}\BibitemShut {NoStop}%
\bibitem [{\citenamefont {Goulko}\ \emph {et~al.}(2016)\citenamefont {Goulko},
  \citenamefont {Mishchenko}, \citenamefont {Prokof'ev},\ and\ \citenamefont
  {Svistunov}}]{Goulko2016PRA}%
  \BibitemOpen
  \bibfield  {author} {\bibinfo {author} {\bibfnamefont {Olga}\ \bibnamefont
  {Goulko}}, \bibinfo {author} {\bibfnamefont {Andrey~S.}\ \bibnamefont
  {Mishchenko}}, \bibinfo {author} {\bibfnamefont {Nikolay}\ \bibnamefont
  {Prokof'ev}}, \ and\ \bibinfo {author} {\bibfnamefont {Boris}\ \bibnamefont
  {Svistunov}},\ }\bibfield  {title} {\enquote {\bibinfo {title} {Dark
  continuum in the spectral function of the resonant fermi polaron},}\
  }\href@noop {} {\bibfield  {journal} {\bibinfo  {journal} {Phys. Rev. A}\
  }\textbf {\bibinfo {volume} {94}},\ \bibinfo {pages} {051605} (\bibinfo
  {year} {2016})}\BibitemShut {NoStop}%
\bibitem [{\citenamefont {Pessoa}\ \emph {et~al.}(2021)\citenamefont {Pessoa},
  \citenamefont {Vitiello},\ and\ \citenamefont {Ardila}}]{Pessoa2021PRA}%
  \BibitemOpen
  \bibfield  {author} {\bibinfo {author} {\bibfnamefont {Renato}\ \bibnamefont
  {Pessoa}}, \bibinfo {author} {\bibfnamefont {S.~A.}\ \bibnamefont
  {Vitiello}}, \ and\ \bibinfo {author} {\bibfnamefont {L.~A. Pe\~na}\
  \bibnamefont {Ardila}},\ }\bibfield  {title} {\enquote {\bibinfo {title}
  {Finite-range effects in the unitary fermi polaron},}\ }\href {\doibase
  10.1103/PhysRevA.104.043313} {\bibfield  {journal} {\bibinfo  {journal}
  {Phys. Rev. A}\ }\textbf {\bibinfo {volume} {104}},\ \bibinfo {pages}
  {043313} (\bibinfo {year} {2021})}\BibitemShut {NoStop}%
\bibitem [{\citenamefont {Cetina}\ \emph {et~al.}(2016)\citenamefont {Cetina},
  \citenamefont {Jag}, \citenamefont {Lous}, \citenamefont {Fritsche},
  \citenamefont {M.Walraven}, \citenamefont {Grimm}, \citenamefont {Levinsen},
  \citenamefont {Parish}, \citenamefont {Schmidt}, \citenamefont {Knap},\ and\
  \citenamefont {Demler}}]{Demler2016Science}%
  \BibitemOpen
  \bibfield  {author} {\bibinfo {author} {\bibfnamefont {M.}~\bibnamefont
  {Cetina}}, \bibinfo {author} {\bibfnamefont {M.}~\bibnamefont {Jag}},
  \bibinfo {author} {\bibfnamefont {R.~S.}\ \bibnamefont {Lous}}, \bibinfo
  {author} {\bibfnamefont {I.}~\bibnamefont {Fritsche}}, \bibinfo {author}
  {\bibfnamefont {J.~T.}\ \bibnamefont {M.Walraven}}, \bibinfo {author}
  {\bibfnamefont {R.}~\bibnamefont {Grimm}}, \bibinfo {author} {\bibfnamefont
  {J.}~\bibnamefont {Levinsen}}, \bibinfo {author} {\bibfnamefont {M.~M.}\
  \bibnamefont {Parish}}, \bibinfo {author} {\bibfnamefont {R.}~\bibnamefont
  {Schmidt}}, \bibinfo {author} {\bibfnamefont {M.}~\bibnamefont {Knap}}, \
  and\ \bibinfo {author} {\bibfnamefont {E.}~\bibnamefont {Demler}},\
  }\bibfield  {title} {\enquote {\bibinfo {title} {Ultrafast many-body
  interferometry of impurities coupled to a fermi sea},}\ }\href@noop {}
  {\bibfield  {journal} {\bibinfo  {journal} {Science}\ }\textbf {\bibinfo
  {volume} {354}},\ \bibinfo {pages} {96} (\bibinfo {year} {2016})}\BibitemShut
  {NoStop}%
\bibitem [{\citenamefont {Liu}\ \emph {et~al.}(2019)\citenamefont {Liu},
  \citenamefont {Levinsen},\ and\ \citenamefont {Parish}}]{Meera2019PRL}%
  \BibitemOpen
  \bibfield  {author} {\bibinfo {author} {\bibfnamefont {Weizhe~Edward}\
  \bibnamefont {Liu}}, \bibinfo {author} {\bibfnamefont {Jesper}\ \bibnamefont
  {Levinsen}}, \ and\ \bibinfo {author} {\bibfnamefont {Meera~M.}\ \bibnamefont
  {Parish}},\ }\bibfield  {title} {\enquote {\bibinfo {title} {Variational
  approach for impurity dynamics at finite temperature},}\ }\href@noop {}
  {\bibfield  {journal} {\bibinfo  {journal} {Phys. Rev. Lett.}\ }\textbf
  {\bibinfo {volume} {122}},\ \bibinfo {pages} {205301} (\bibinfo {year}
  {2019})}\BibitemShut {NoStop}%
\bibitem [{\citenamefont {You}\ \emph {et~al.}(2019)\citenamefont {You},
  \citenamefont {Schmidt}, \citenamefont {Ivanov}, \citenamefont {Knap},\ and\
  \citenamefont {Demler}}]{Demler2019PRB}%
  \BibitemOpen
  \bibfield  {author} {\bibinfo {author} {\bibfnamefont {Jhih-Shih}\
  \bibnamefont {You}}, \bibinfo {author} {\bibfnamefont {Richard}\ \bibnamefont
  {Schmidt}}, \bibinfo {author} {\bibfnamefont {Dmitri~A.}\ \bibnamefont
  {Ivanov}}, \bibinfo {author} {\bibfnamefont {Michael}\ \bibnamefont {Knap}},
  \ and\ \bibinfo {author} {\bibfnamefont {Eugene}\ \bibnamefont {Demler}},\
  }\bibfield  {title} {\enquote {\bibinfo {title} {Atomtronics with a spin:
  Statistics of spin transport and nonequilibrium orthogonality catastrophe in
  cold quantum gases},}\ }\href@noop {} {\bibfield  {journal} {\bibinfo
  {journal} {Phys. Rev. B}\ }\textbf {\bibinfo {volume} {99}},\ \bibinfo
  {pages} {214505} (\bibinfo {year} {2019})}\BibitemShut {NoStop}%
\bibitem [{\citenamefont {Mitchison}\ \emph {et~al.}(2020)\citenamefont
  {Mitchison}, \citenamefont {Fogarty}, \citenamefont {Guarnieri},
  \citenamefont {Campbell}, \citenamefont {Busch},\ and\ \citenamefont
  {Goold}}]{Goold2020PRL}%
  \BibitemOpen
  \bibfield  {author} {\bibinfo {author} {\bibfnamefont {Mark~T.}\ \bibnamefont
  {Mitchison}}, \bibinfo {author} {\bibfnamefont {Thom\'as}\ \bibnamefont
  {Fogarty}}, \bibinfo {author} {\bibfnamefont {Giacomo}\ \bibnamefont
  {Guarnieri}}, \bibinfo {author} {\bibfnamefont {Steve}\ \bibnamefont
  {Campbell}}, \bibinfo {author} {\bibfnamefont {Thomas}\ \bibnamefont
  {Busch}}, \ and\ \bibinfo {author} {\bibfnamefont {John}\ \bibnamefont
  {Goold}},\ }\bibfield  {title} {\enquote {\bibinfo {title} {In situ
  thermometry of a cold fermi gas via dephasing impurities},}\ }\href@noop {}
  {\bibfield  {journal} {\bibinfo  {journal} {Phys. Rev. Lett.}\ }\textbf
  {\bibinfo {volume} {125}},\ \bibinfo {pages} {080402} (\bibinfo {year}
  {2020})}\BibitemShut {NoStop}%
\bibitem [{\citenamefont {Braaten}\ \emph {et~al.}(2010)\citenamefont
  {Braaten}, \citenamefont {Kang},\ and\ \citenamefont
  {Platter}}]{Braaten2010PRL}%
  \BibitemOpen
  \bibfield  {author} {\bibinfo {author} {\bibfnamefont {Eric}\ \bibnamefont
  {Braaten}}, \bibinfo {author} {\bibfnamefont {Daekyoung}\ \bibnamefont
  {Kang}}, \ and\ \bibinfo {author} {\bibfnamefont {Lucas}\ \bibnamefont
  {Platter}},\ }\bibfield  {title} {\enquote {\bibinfo {title} {Short-time
  operator product expansion for rf spectroscopy of a strongly interacting
  fermi gas},}\ }\href@noop {} {\bibfield  {journal} {\bibinfo  {journal}
  {Phys. Rev. Lett.}\ }\textbf {\bibinfo {volume} {104}},\ \bibinfo {pages}
  {223004} (\bibinfo {year} {2010})}\BibitemShut {NoStop}%
\bibitem [{\citenamefont {Liu}\ \emph {et~al.}(2020)\citenamefont {Liu},
  \citenamefont {Shi}, \citenamefont {Parish},\ and\ \citenamefont
  {Levinsen}}]{Meera2020PRA}%
  \BibitemOpen
  \bibfield  {author} {\bibinfo {author} {\bibfnamefont {Weizhe~Edward}\
  \bibnamefont {Liu}}, \bibinfo {author} {\bibfnamefont {Zhe-Yu}\ \bibnamefont
  {Shi}}, \bibinfo {author} {\bibfnamefont {Meera~M.}\ \bibnamefont {Parish}},
  \ and\ \bibinfo {author} {\bibfnamefont {Jesper}\ \bibnamefont {Levinsen}},\
  }\bibfield  {title} {\enquote {\bibinfo {title} {Theory of radio-frequency
  spectroscopy of impurities in quantum gases},}\ }\href@noop {} {\bibfield
  {journal} {\bibinfo  {journal} {Phys. Rev. A}\ }\textbf {\bibinfo {volume}
  {102}},\ \bibinfo {pages} {023304} (\bibinfo {year} {2020})}\BibitemShut
  {NoStop}%
\bibitem [{\citenamefont {Adlong}\ \emph {et~al.}(2021)\citenamefont {Adlong},
  \citenamefont {Liu}, \citenamefont {Turner}, \citenamefont {Parish},\ and\
  \citenamefont {Levinsen}}]{Meera2021PRA}%
  \BibitemOpen
  \bibfield  {author} {\bibinfo {author} {\bibfnamefont {Haydn~S.}\
  \bibnamefont {Adlong}}, \bibinfo {author} {\bibfnamefont {Weizhe~Edward}\
  \bibnamefont {Liu}}, \bibinfo {author} {\bibfnamefont {Lincoln~D.}\
  \bibnamefont {Turner}}, \bibinfo {author} {\bibfnamefont {Meera~M.}\
  \bibnamefont {Parish}}, \ and\ \bibinfo {author} {\bibfnamefont {Jesper}\
  \bibnamefont {Levinsen}},\ }\bibfield  {title} {\enquote {\bibinfo {title}
  {Signatures of the orthogonality catastrophe in a coherently driven
  impurity},}\ }\href@noop {} {\bibfield  {journal} {\bibinfo  {journal} {Phys.
  Rev. A}\ }\textbf {\bibinfo {volume} {104}},\ \bibinfo {pages} {043309}
  (\bibinfo {year} {2021})}\BibitemShut {NoStop}%
\bibitem [{\citenamefont {Balewski}\ \emph {et~al.}(2013)\citenamefont
  {Balewski}, \citenamefont {Krupp}, \citenamefont {Gaj}, \citenamefont
  {Peter}, \citenamefont {B{'"u}chler}, \citenamefont {L{\"o}w}, \citenamefont
  {Hofferberth},\ and\ \citenamefont {Pfau}}]{Pfau2013Nature}%
  \BibitemOpen
  \bibfield  {author} {\bibinfo {author} {\bibfnamefont {Jonathan~B.}\
  \bibnamefont {Balewski}}, \bibinfo {author} {\bibfnamefont {Alexander~T.}\
  \bibnamefont {Krupp}}, \bibinfo {author} {\bibfnamefont {Anita}\ \bibnamefont
  {Gaj}}, \bibinfo {author} {\bibfnamefont {David}\ \bibnamefont {Peter}},
  \bibinfo {author} {\bibfnamefont {Hans~Peter}\ \bibnamefont {B{'"u}chler}},
  \bibinfo {author} {\bibfnamefont {Robert}\ \bibnamefont {L{\"o}w}}, \bibinfo
  {author} {\bibfnamefont {Sebastian}\ \bibnamefont {Hofferberth}}, \ and\
  \bibinfo {author} {\bibfnamefont {Tilman}\ \bibnamefont {Pfau}},\ }\bibfield
  {title} {\enquote {\bibinfo {title} {Coupling a single electron to a
  bose-einstein condensate},}\ }\href@noop {} {\bibfield  {journal} {\bibinfo
  {journal} {Nature (London)}\ }\textbf {\bibinfo {volume} {502}},\ \bibinfo
  {pages} {664--667} (\bibinfo {year} {2013})}\BibitemShut {NoStop}%
\bibitem [{\citenamefont {Wang}\ \emph {et~al.}(2015)\citenamefont {Wang},
  \citenamefont {Gacesa},\ and\ \citenamefont {C{\^o}t{\'e}}}]{Jia2015PRL}%
  \BibitemOpen
  \bibfield  {author} {\bibinfo {author} {\bibfnamefont {Jia}\ \bibnamefont
  {Wang}}, \bibinfo {author} {\bibfnamefont {Marko}\ \bibnamefont {Gacesa}}, \
  and\ \bibinfo {author} {\bibfnamefont {R.}~\bibnamefont {C{\^o}t{\'e}}},\
  }\bibfield  {title} {\enquote {\bibinfo {title} {Rydberg electrons in a
  {Bose-Einstein} condensate},}\ }\href@noop {} {\bibfield  {journal} {\bibinfo
   {journal} {Phys. Rev. Lett.}\ }\textbf {\bibinfo {volume} {114}},\ \bibinfo
  {pages} {243003} (\bibinfo {year} {2015})}\BibitemShut {NoStop}%
\bibitem [{\citenamefont {Sous}\ \emph {et~al.}(2020)\citenamefont {Sous},
  \citenamefont {Sadeghpour}, \citenamefont {Killian}, \citenamefont {Demler},\
  and\ \citenamefont {Schmidt}}]{Schmidt2020PRR}%
  \BibitemOpen
  \bibfield  {author} {\bibinfo {author} {\bibfnamefont {John}\ \bibnamefont
  {Sous}}, \bibinfo {author} {\bibfnamefont {H.~R.}\ \bibnamefont
  {Sadeghpour}}, \bibinfo {author} {\bibfnamefont {T.~C.}\ \bibnamefont
  {Killian}}, \bibinfo {author} {\bibfnamefont {Eugene}\ \bibnamefont
  {Demler}}, \ and\ \bibinfo {author} {\bibfnamefont {Richard}\ \bibnamefont
  {Schmidt}},\ }\bibfield  {title} {\enquote {\bibinfo {title} {Rydberg
  impurity in a fermi gas: Quantum statistics and rotational blockade},}\
  }\href@noop {} {\bibfield  {journal} {\bibinfo  {journal} {Phys. Rev.
  Research}\ }\textbf {\bibinfo {volume} {2}},\ \bibinfo {pages} {023021}
  (\bibinfo {year} {2020})}\BibitemShut {NoStop}%
\bibitem [{\citenamefont {Schmidt}\ \emph {et~al.}(2016)\citenamefont
  {Schmidt}, \citenamefont {Sadeghpour},\ and\ \citenamefont
  {Demler}}]{Schmidt2016PRL}%
  \BibitemOpen
  \bibfield  {author} {\bibinfo {author} {\bibfnamefont {Richard}\ \bibnamefont
  {Schmidt}}, \bibinfo {author} {\bibfnamefont {H.~R.}\ \bibnamefont
  {Sadeghpour}}, \ and\ \bibinfo {author} {\bibfnamefont {E.}~\bibnamefont
  {Demler}},\ }\bibfield  {title} {\enquote {\bibinfo {title} {Mesoscopic
  rydberg impurity in an atomic quantum gas},}\ }\href@noop {} {\bibfield
  {journal} {\bibinfo  {journal} {Phys. Rev. Lett.}\ }\textbf {\bibinfo
  {volume} {116}},\ \bibinfo {pages} {105302} (\bibinfo {year}
  {2016})}\BibitemShut {NoStop}%
\bibitem [{\citenamefont {Camargo}\ \emph {et~al.}(2018)\citenamefont
  {Camargo}, \citenamefont {Schmidt}, \citenamefont {Whalen}, \citenamefont
  {Ding}, \citenamefont {Woehl}, \citenamefont {Yoshida}, \citenamefont
  {Burgd\"orfer}, \citenamefont {Dunning}, \citenamefont {Sadeghpour},
  \citenamefont {Demler},\ and\ \citenamefont {Killian}}]{Camargo2018PRL}%
  \BibitemOpen
  \bibfield  {author} {\bibinfo {author} {\bibfnamefont {F.}~\bibnamefont
  {Camargo}}, \bibinfo {author} {\bibfnamefont {R.}~\bibnamefont {Schmidt}},
  \bibinfo {author} {\bibfnamefont {J.~D.}\ \bibnamefont {Whalen}}, \bibinfo
  {author} {\bibfnamefont {R.}~\bibnamefont {Ding}}, \bibinfo {author}
  {\bibfnamefont {G.}~\bibnamefont {Woehl}}, \bibinfo {author} {\bibfnamefont
  {S.}~\bibnamefont {Yoshida}}, \bibinfo {author} {\bibfnamefont
  {J.}~\bibnamefont {Burgd\"orfer}}, \bibinfo {author} {\bibfnamefont {F.~B.}\
  \bibnamefont {Dunning}}, \bibinfo {author} {\bibfnamefont {H.~R.}\
  \bibnamefont {Sadeghpour}}, \bibinfo {author} {\bibfnamefont
  {E.}~\bibnamefont {Demler}}, \ and\ \bibinfo {author} {\bibfnamefont {T.~C.}\
  \bibnamefont {Killian}},\ }\bibfield  {title} {\enquote {\bibinfo {title}
  {Creation of rydberg polarons in a bose gas},}\ }\href@noop {} {\bibfield
  {journal} {\bibinfo  {journal} {Phys. Rev. Lett.}\ }\textbf {\bibinfo
  {volume} {120}},\ \bibinfo {pages} {083401} (\bibinfo {year}
  {2018})}\BibitemShut {NoStop}%
\bibitem [{\citenamefont {Schmidt}\ \emph
  {et~al.}(2018{\natexlab{b}})\citenamefont {Schmidt}, \citenamefont {Whalen},
  \citenamefont {Ding}, \citenamefont {Camargo}, \citenamefont {Woehl},
  \citenamefont {Yoshida}, \citenamefont {Burgd\"orfer}, \citenamefont
  {Dunning}, \citenamefont {Demler}, \citenamefont {Sadeghpour},\ and\
  \citenamefont {Killian}}]{Schmidt2018PRA}%
  \BibitemOpen
  \bibfield  {author} {\bibinfo {author} {\bibfnamefont {R.}~\bibnamefont
  {Schmidt}}, \bibinfo {author} {\bibfnamefont {J.~D.}\ \bibnamefont {Whalen}},
  \bibinfo {author} {\bibfnamefont {R.}~\bibnamefont {Ding}}, \bibinfo {author}
  {\bibfnamefont {F.}~\bibnamefont {Camargo}}, \bibinfo {author} {\bibfnamefont
  {G.}~\bibnamefont {Woehl}}, \bibinfo {author} {\bibfnamefont
  {S.}~\bibnamefont {Yoshida}}, \bibinfo {author} {\bibfnamefont
  {J.}~\bibnamefont {Burgd\"orfer}}, \bibinfo {author} {\bibfnamefont {F.~B.}\
  \bibnamefont {Dunning}}, \bibinfo {author} {\bibfnamefont {E.}~\bibnamefont
  {Demler}}, \bibinfo {author} {\bibfnamefont {H.~R.}\ \bibnamefont
  {Sadeghpour}}, \ and\ \bibinfo {author} {\bibfnamefont {T.~C.}\ \bibnamefont
  {Killian}},\ }\bibfield  {title} {\enquote {\bibinfo {title} {Theory of
  excitation of rydberg polarons in an atomic quantum gas},}\ }\href@noop {}
  {\bibfield  {journal} {\bibinfo  {journal} {Phys. Rev. A}\ }\textbf {\bibinfo
  {volume} {97}},\ \bibinfo {pages} {022707} (\bibinfo {year}
  {2018}{\natexlab{b}})}\BibitemShut {NoStop}%
\bibitem [{\citenamefont {Wang}\ \emph
  {et~al.}(2022{\natexlab{a}})\citenamefont {Wang}, \citenamefont {Liu},\ and\
  \citenamefont {Hu}}]{JiaWang2022PRLshort}%
  \BibitemOpen
  \bibfield  {author} {\bibinfo {author} {\bibfnamefont {Jia}\ \bibnamefont
  {Wang}}, \bibinfo {author} {\bibfnamefont {Xia-Ji}\ \bibnamefont {Liu}}, \
  and\ \bibinfo {author} {\bibfnamefont {Hui}\ \bibnamefont {Hu}},\ }\bibfield
  {title} {\enquote {\bibinfo {title} {Exact quasiparticle properties of a
  heavy polaron in bcs fermi superfluids},}\ }\href@noop {} {\bibfield
  {journal} {\bibinfo  {journal} {Phys. Rev. Lett.}\ }\textbf {\bibinfo
  {volume} {128}},\ \bibinfo {pages} {175301} (\bibinfo {year}
  {2022}{\natexlab{a}})}\BibitemShut {NoStop}%
\bibitem [{\citenamefont {Wang}\ \emph
  {et~al.}(2022{\natexlab{b}})\citenamefont {Wang}, \citenamefont {Liu},\ and\
  \citenamefont {Hu}}]{JiaWang2022PRAlong}%
  \BibitemOpen
  \bibfield  {author} {\bibinfo {author} {\bibfnamefont {Jia}\ \bibnamefont
  {Wang}}, \bibinfo {author} {\bibfnamefont {Xia-Ji}\ \bibnamefont {Liu}}, \
  and\ \bibinfo {author} {\bibfnamefont {Hui}\ \bibnamefont {Hu}},\ }\bibfield
  {title} {\enquote {\bibinfo {title} {Heavy polarons in ultracold atomic fermi
  superfluids at the bec-bcs crossover: Formalism and applications},}\
  }\href@noop {} {\bibfield  {journal} {\bibinfo  {journal} {Phys. Rev. A}\
  }\textbf {\bibinfo {volume} {105}},\ \bibinfo {pages} {043320} (\bibinfo
  {year} {2022}{\natexlab{b}})}\BibitemShut {NoStop}%
\bibitem [{\citenamefont {Heyl}(2018)}]{Heyl2018Rpp}%
  \BibitemOpen
  \bibfield  {author} {\bibinfo {author} {\bibfnamefont {Markus}\ \bibnamefont
  {Heyl}},\ }\bibfield  {title} {\enquote {\bibinfo {title} {Dynamical quantum
  phase transitions: a review},}\ }\href@noop {} {\bibfield  {journal}
  {\bibinfo  {journal} {Rep. Prog. Phys}\ }\textbf {\bibinfo {volume} {81}},\
  \bibinfo {pages} {054001} (\bibinfo {year} {2018})}\BibitemShut {NoStop}%
\bibitem [{\citenamefont {Wang}\ \emph {et~al.}(2019)\citenamefont {Wang},
  \citenamefont {Liu},\ and\ \citenamefont {Hu}}]{Jia2019PRL}%
  \BibitemOpen
  \bibfield  {author} {\bibinfo {author} {\bibfnamefont {Jia}\ \bibnamefont
  {Wang}}, \bibinfo {author} {\bibfnamefont {Xia-Ji}\ \bibnamefont {Liu}}, \
  and\ \bibinfo {author} {\bibfnamefont {Hui}\ \bibnamefont {Hu}},\ }\bibfield
  {title} {\enquote {\bibinfo {title} {Roton-induced bose polaron in the
  presence of synthetic spin-orbit coupling},}\ }\href@noop {} {\bibfield
  {journal} {\bibinfo  {journal} {Phys. Rev. Lett.}\ }\textbf {\bibinfo
  {volume} {123}},\ \bibinfo {pages} {213401} (\bibinfo {year}
  {2019})}\BibitemShut {NoStop}%
\bibitem [{\citenamefont {Nishida}(2015)}]{Nishida2015PRL}%
  \BibitemOpen
  \bibfield  {author} {\bibinfo {author} {\bibfnamefont {Yusuke}\ \bibnamefont
  {Nishida}},\ }\bibfield  {title} {\enquote {\bibinfo {title} {Polaronic
  atom-trimer continuity in three-component fermi gases},}\ }\href@noop {}
  {\bibfield  {journal} {\bibinfo  {journal} {Phys. Rev. Lett.}\ }\textbf
  {\bibinfo {volume} {114}},\ \bibinfo {pages} {115302} (\bibinfo {year}
  {2015})}\BibitemShut {NoStop}%
\bibitem [{\citenamefont {Yi}\ and\ \citenamefont {Cui}(2015)}]{Yi2015PRA}%
  \BibitemOpen
  \bibfield  {author} {\bibinfo {author} {\bibfnamefont {Wei}\ \bibnamefont
  {Yi}}\ and\ \bibinfo {author} {\bibfnamefont {Xiaoling}\ \bibnamefont
  {Cui}},\ }\bibfield  {title} {\enquote {\bibinfo {title} {Polarons in
  ultracold fermi superfluids},}\ }\href@noop {} {\bibfield  {journal}
  {\bibinfo  {journal} {Phys. Rev. A}\ }\textbf {\bibinfo {volume} {92}},\
  \bibinfo {pages} {013620} (\bibinfo {year} {2015})}\BibitemShut {NoStop}%
\bibitem [{\citenamefont {Pierce}\ \emph {et~al.}(2019)\citenamefont {Pierce},
  \citenamefont {Leyronas},\ and\ \citenamefont {Chevy}}]{Pierce2019PRL}%
  \BibitemOpen
  \bibfield  {author} {\bibinfo {author} {\bibfnamefont {M.}~\bibnamefont
  {Pierce}}, \bibinfo {author} {\bibfnamefont {X.}~\bibnamefont {Leyronas}}, \
  and\ \bibinfo {author} {\bibfnamefont {F.}~\bibnamefont {Chevy}},\ }\bibfield
   {title} {\enquote {\bibinfo {title} {Few versus many-body physics of an
  impurity immersed in a superfluid of spin $1/2$ attractive fermions},}\
  }\href@noop {} {\bibfield  {journal} {\bibinfo  {journal} {Phys. Rev. Lett.}\
  }\textbf {\bibinfo {volume} {123}},\ \bibinfo {pages} {080403} (\bibinfo
  {year} {2019})}\BibitemShut {NoStop}%
\bibitem [{\citenamefont {Hu}\ \emph {et~al.}(2022{\natexlab{a}})\citenamefont
  {Hu}, \citenamefont {Wang}, \citenamefont {Zhou},\ and\ \citenamefont
  {Liu}}]{HuiHu2022PRA1}%
  \BibitemOpen
  \bibfield  {author} {\bibinfo {author} {\bibfnamefont {Hui}\ \bibnamefont
  {Hu}}, \bibinfo {author} {\bibfnamefont {Jia}\ \bibnamefont {Wang}}, \bibinfo
  {author} {\bibfnamefont {Jing}\ \bibnamefont {Zhou}}, \ and\ \bibinfo
  {author} {\bibfnamefont {Xia-Ji}\ \bibnamefont {Liu}},\ }\bibfield  {title}
  {\enquote {\bibinfo {title} {Crossover polarons in a strongly interacting
  fermi superfluid},}\ }\href@noop {} {\bibfield  {journal} {\bibinfo
  {journal} {Phys. Rev. A}\ }\textbf {\bibinfo {volume} {105}},\ \bibinfo
  {pages} {023317} (\bibinfo {year} {2022}{\natexlab{a}})}\BibitemShut
  {NoStop}%
\bibitem [{\citenamefont {Bigu\'e}\ \emph {et~al.}(2022)\citenamefont
  {Bigu\'e}, \citenamefont {Chevy},\ and\ \citenamefont
  {Leyronas}}]{Bigue2022PRA}%
  \BibitemOpen
  \bibfield  {author} {\bibinfo {author} {\bibfnamefont {A.}~\bibnamefont
  {Bigu\'e}}, \bibinfo {author} {\bibfnamefont {F.}~\bibnamefont {Chevy}}, \
  and\ \bibinfo {author} {\bibfnamefont {X.}~\bibnamefont {Leyronas}},\
  }\bibfield  {title} {\enquote {\bibinfo {title} {Mean field versus
  random-phase approximation calculation of the energy of an impurity immersed
  in a spin-1/2 superfluid},}\ }\href@noop {} {\bibfield  {journal} {\bibinfo
  {journal} {Phys. Rev. A}\ }\textbf {\bibinfo {volume} {105}},\ \bibinfo
  {pages} {033314} (\bibinfo {year} {2022})}\BibitemShut {NoStop}%
\bibitem [{\citenamefont {Yu}(1965)}]{Yu1965ActaPhysSin}%
  \BibitemOpen
  \bibfield  {author} {\bibinfo {author} {\bibfnamefont {L.}~\bibnamefont
  {Yu}},\ }\bibfield  {title} {\enquote {\bibinfo {title} {Bound state in
  superconductors with paramagnetic impurities},}\ }\href@noop {} {\bibfield
  {journal} {\bibinfo  {journal} {Acta. Phys. Sin.}\ }\textbf {\bibinfo
  {volume} {21}},\ \bibinfo {pages} {75} (\bibinfo {year} {1965})}\BibitemShut
  {NoStop}%
\bibitem [{\citenamefont {Shiba}(1968)}]{Shiba1968ProgTheorPhys}%
  \BibitemOpen
  \bibfield  {author} {\bibinfo {author} {\bibfnamefont {H.}~\bibnamefont
  {Shiba}},\ }\bibfield  {title} {\enquote {\bibinfo {title} {Classical spin in
  superconductors},}\ }\href@noop {} {\bibfield  {journal} {\bibinfo  {journal}
  {Prog. Theor. Phys.}\ }\textbf {\bibinfo {volume} {40}},\ \bibinfo {pages}
  {435} (\bibinfo {year} {1968})}\BibitemShut {NoStop}%
\bibitem [{\citenamefont {Rusinov}(1969)}]{Rusinov1969JETP}%
  \BibitemOpen
  \bibfield  {author} {\bibinfo {author} {\bibfnamefont {A.~I.}\ \bibnamefont
  {Rusinov}},\ }\bibfield  {title} {\enquote {\bibinfo {title}
  {Superconductivity near a paramagnetic impurity},}\ }\href@noop {} {\bibfield
   {journal} {\bibinfo  {journal} {JETP Lett. (USSR)}\ }\textbf {\bibinfo
  {volume} {9}},\ \bibinfo {pages} {85} (\bibinfo {year} {1969})}\BibitemShut
  {NoStop}%
\bibitem [{\citenamefont {Vernier}\ \emph {et~al.}(2011)\citenamefont
  {Vernier}, \citenamefont {Pekker}, \citenamefont {Zwierlein},\ and\
  \citenamefont {Demler}}]{Vernier2011PRA}%
  \BibitemOpen
  \bibfield  {author} {\bibinfo {author} {\bibfnamefont {Eric}\ \bibnamefont
  {Vernier}}, \bibinfo {author} {\bibfnamefont {David}\ \bibnamefont {Pekker}},
  \bibinfo {author} {\bibfnamefont {Martin~W.}\ \bibnamefont {Zwierlein}}, \
  and\ \bibinfo {author} {\bibfnamefont {Eugene}\ \bibnamefont {Demler}},\
  }\bibfield  {title} {\enquote {\bibinfo {title} {Bound states of a localized
  magnetic impurity in a superfluid of paired ultracold fermions},}\
  }\href@noop {} {\bibfield  {journal} {\bibinfo  {journal} {Phys. Rev. A}\
  }\textbf {\bibinfo {volume} {83}},\ \bibinfo {pages} {033619} (\bibinfo
  {year} {2011})}\BibitemShut {NoStop}%
\bibitem [{\citenamefont {Jiang}\ \emph {et~al.}(2011)\citenamefont {Jiang},
  \citenamefont {Baksmaty}, \citenamefont {Hu}, \citenamefont {Chen},\ and\
  \citenamefont {Pu}}]{Jiang2011PRA}%
  \BibitemOpen
  \bibfield  {author} {\bibinfo {author} {\bibfnamefont {Lei}\ \bibnamefont
  {Jiang}}, \bibinfo {author} {\bibfnamefont {Leslie~O.}\ \bibnamefont
  {Baksmaty}}, \bibinfo {author} {\bibfnamefont {Hui}\ \bibnamefont {Hu}},
  \bibinfo {author} {\bibfnamefont {Yan}\ \bibnamefont {Chen}}, \ and\ \bibinfo
  {author} {\bibfnamefont {Han}\ \bibnamefont {Pu}},\ }\bibfield  {title}
  {\enquote {\bibinfo {title} {Single impurity in ultracold fermi
  superfluids},}\ }\href@noop {} {\bibfield  {journal} {\bibinfo  {journal}
  {Phys. Rev. A}\ }\textbf {\bibinfo {volume} {83}},\ \bibinfo {pages} {061604}
  (\bibinfo {year} {2011})}\BibitemShut {NoStop}%
\bibitem [{\citenamefont {Gurarie}\ and\ \citenamefont
  {Radzihovsky}(2007)}]{Gurarie2006AnnPhys}%
  \BibitemOpen
  \bibfield  {author} {\bibinfo {author} {\bibfnamefont {V.}~\bibnamefont
  {Gurarie}}\ and\ \bibinfo {author} {\bibfnamefont {L.}~\bibnamefont
  {Radzihovsky}},\ }\bibfield  {title} {\enquote {\bibinfo {title}
  {Resonantly-paired fermionic superfluids},}\ }\href@noop {} {\bibfield
  {journal} {\bibinfo  {journal} {Ann. Phys. (N. Y.)}\ }\textbf {\bibinfo
  {volume} {332}},\ \bibinfo {pages} {2} (\bibinfo {year} {2007})}\BibitemShut
  {NoStop}%
\bibitem [{\citenamefont {Balatsky}\ \emph {et~al.}(2006)\citenamefont
  {Balatsky}, \citenamefont {Vekhter},\ and\ \citenamefont
  {Zhu}}]{Balatsky2006RMP}%
  \BibitemOpen
  \bibfield  {author} {\bibinfo {author} {\bibfnamefont {A.~V.}\ \bibnamefont
  {Balatsky}}, \bibinfo {author} {\bibfnamefont {I.}~\bibnamefont {Vekhter}}, \
  and\ \bibinfo {author} {\bibfnamefont {Jian-Xin}\ \bibnamefont {Zhu}},\
  }\bibfield  {title} {\enquote {\bibinfo {title} {Impurity-induced states in
  conventional and unconventional superconductors},}\ }\href@noop {} {\bibfield
   {journal} {\bibinfo  {journal} {Rev. Mod. Phys.}\ }\textbf {\bibinfo
  {volume} {78}},\ \bibinfo {pages} {373--433} (\bibinfo {year}
  {2006})}\BibitemShut {NoStop}%
\bibitem [{\citenamefont {Wang}(2022)}]{JiaWang2022arXiv1}%
  \BibitemOpen
  \bibfield  {author} {\bibinfo {author} {\bibfnamefont {Jia}\ \bibnamefont
  {Wang}},\ }\href@noop {} {\enquote {\bibinfo {title} {Multidimensional
  spectroscopy of time-dependent impurities in ultracold fermions},}\ }
  (\bibinfo {year} {2022}),\ \bibinfo {note} {{arXiv}:2207.10501}\BibitemShut
  {NoStop}%
\bibitem [{\citenamefont {Tempelaar}\ and\ \citenamefont
  {Berkelbach}(2019)}]{Tempelaa2019NC}%
  \BibitemOpen
  \bibfield  {author} {\bibinfo {author} {\bibfnamefont {Roel}\ \bibnamefont
  {Tempelaar}}\ and\ \bibinfo {author} {\bibfnamefont {Timothy~C.}\
  \bibnamefont {Berkelbach}},\ }\bibfield  {title} {\enquote {\bibinfo {title}
  {Many-body simulation of two-dimensional electronic spectroscopy of excitons
  and trions in monolayer transition metal dichalcogenides},}\ }\href@noop {}
  {\bibfield  {journal} {\bibinfo  {journal} {Nat. Commun.}\ }\textbf {\bibinfo
  {volume} {10}},\ \bibinfo {pages} {3419} (\bibinfo {year}
  {2019})}\BibitemShut {NoStop}%
\bibitem [{\citenamefont {Hu}\ \emph {et~al.}(2022{\natexlab{b}})\citenamefont
  {Hu}, \citenamefont {Wang},\ and\ \citenamefont {Liu}}]{HuiHu2022arXiv}%
  \BibitemOpen
  \bibfield  {author} {\bibinfo {author} {\bibfnamefont {Hui}\ \bibnamefont
  {Hu}}, \bibinfo {author} {\bibfnamefont {Jia}\ \bibnamefont {Wang}}, \ and\
  \bibinfo {author} {\bibfnamefont {Xia-Ji}\ \bibnamefont {Liu}},\ }\href@noop
  {} {\enquote {\bibinfo {title} {Microscopic many-body theory of
  two-dimensional coherent spectroscopy of excitons and trions in atomically
  thin transition metal dichalcogenides},}\ } (\bibinfo {year}
  {2022}{\natexlab{b}}),\ \bibinfo {note} {{arXiv}:2208.03599}\BibitemShut
  {NoStop}%
\bibitem [{\citenamefont {Nardin}\ \emph {et~al.}(2015)\citenamefont {Nardin},
  \citenamefont {Autry}, \citenamefont {Moody}, \citenamefont {Singh},
  \citenamefont {Li},\ and\ \citenamefont {Cundiff}}]{Cundiff2015JApplP}%
  \BibitemOpen
  \bibfield  {author} {\bibinfo {author} {\bibfnamefont {Ga{\"e}l}\
  \bibnamefont {Nardin}}, \bibinfo {author} {\bibfnamefont {Travis~M.}\
  \bibnamefont {Autry}}, \bibinfo {author} {\bibfnamefont {Galan}\ \bibnamefont
  {Moody}}, \bibinfo {author} {\bibfnamefont {Rohan}\ \bibnamefont {Singh}},
  \bibinfo {author} {\bibfnamefont {Hebin}\ \bibnamefont {Li}}, \ and\ \bibinfo
  {author} {\bibfnamefont {Steven~T.}\ \bibnamefont {Cundiff}},\ }\bibfield
  {title} {\enquote {\bibinfo {title} {Multi-dimensional coherent optical
  spectroscopy of semiconductor nanostructures: Collinear and non-collinear
  approaches},}\ }\href@noop {} {\bibfield  {journal} {\bibinfo  {journal} {J.
  Appl. Phys}\ }\textbf {\bibinfo {volume} {177}},\ \bibinfo {pages} {112804}
  (\bibinfo {year} {2015})}\BibitemShut {NoStop}%
\bibitem [{\citenamefont {Hao}\ \emph {et~al.}(2016)\citenamefont {Hao},
  \citenamefont {Xu}, \citenamefont {Nagler}, \citenamefont {Singh},
  \citenamefont {Tran}, \citenamefont {Dass}, \citenamefont {Sch{\"u}ller},
  \citenamefont {Korn}, \citenamefont {Li},\ and\ \citenamefont
  {Moody}}]{XiaoqinLi2016NL}%
  \BibitemOpen
  \bibfield  {author} {\bibinfo {author} {\bibfnamefont {Kai}\ \bibnamefont
  {Hao}}, \bibinfo {author} {\bibfnamefont {Lixiang}\ \bibnamefont {Xu}},
  \bibinfo {author} {\bibfnamefont {Philipp}\ \bibnamefont {Nagler}}, \bibinfo
  {author} {\bibfnamefont {Akshay}\ \bibnamefont {Singh}}, \bibinfo {author}
  {\bibfnamefont {Kha}\ \bibnamefont {Tran}}, \bibinfo {author} {\bibfnamefont
  {Chandriker~Kavir}\ \bibnamefont {Dass}}, \bibinfo {author} {\bibfnamefont
  {Christian}\ \bibnamefont {Sch{\"u}ller}}, \bibinfo {author} {\bibfnamefont
  {Tobias}\ \bibnamefont {Korn}}, \bibinfo {author} {\bibfnamefont {Xiaoqin}\
  \bibnamefont {Li}}, \ and\ \bibinfo {author} {\bibfnamefont {Galan}\
  \bibnamefont {Moody}},\ }\bibfield  {title} {\enquote {\bibinfo {title}
  {Coherent and incoherent coupling dynamics between neutral and charged
  excitons in monolayer mose{$_2$}},}\ }\href@noop {} {\bibfield  {journal}
  {\bibinfo  {journal} {Nano Lett.}\ }\textbf {\bibinfo {volume} {16}},\
  \bibinfo {pages} {5109} (\bibinfo {year} {2016})}\BibitemShut {NoStop}%
\bibitem [{\citenamefont {Hao}\ \emph {et~al.}(2017)\citenamefont {Hao},
  \citenamefont {Specht}, \citenamefont {Nagler}, \citenamefont {Xu},
  \citenamefont {Tran}, \citenamefont {Singh}, \citenamefont {Dass},
  \citenamefont {Sch{\"u}ller}, \citenamefont {Korn}, \citenamefont {Richter},
  \citenamefont {Knorr}, \citenamefont {Li},\ and\ \citenamefont
  {Moody}}]{XiaoqinLi2017NC}%
  \BibitemOpen
  \bibfield  {author} {\bibinfo {author} {\bibfnamefont {Kai}\ \bibnamefont
  {Hao}}, \bibinfo {author} {\bibfnamefont {Judith~F.}\ \bibnamefont {Specht}},
  \bibinfo {author} {\bibfnamefont {Philipp}\ \bibnamefont {Nagler}}, \bibinfo
  {author} {\bibfnamefont {Lixiang}\ \bibnamefont {Xu}}, \bibinfo {author}
  {\bibfnamefont {Kha}\ \bibnamefont {Tran}}, \bibinfo {author} {\bibfnamefont
  {Akshay}\ \bibnamefont {Singh}}, \bibinfo {author} {\bibfnamefont
  {Chandriker~Kavir}\ \bibnamefont {Dass}}, \bibinfo {author} {\bibfnamefont
  {Christian}\ \bibnamefont {Sch{\"u}ller}}, \bibinfo {author} {\bibfnamefont
  {Tobias}\ \bibnamefont {Korn}}, \bibinfo {author} {\bibfnamefont {Marten}\
  \bibnamefont {Richter}}, \bibinfo {author} {\bibfnamefont {Andreas}\
  \bibnamefont {Knorr}}, \bibinfo {author} {\bibfnamefont {Xiaoqin}\
  \bibnamefont {Li}}, \ and\ \bibinfo {author} {\bibfnamefont {Galan}\
  \bibnamefont {Moody}},\ }\bibfield  {title} {\enquote {\bibinfo {title}
  {Neutral and charged inter-valley biexcitons in monolayer mose{$_2$}},}\
  }\href@noop {} {\bibfield  {journal} {\bibinfo  {journal} {Nat. Commun.}\
  }\textbf {\bibinfo {volume} {8}},\ \bibinfo {pages} {15552} (\bibinfo {year}
  {2017})}\BibitemShut {NoStop}%
\bibitem [{\citenamefont {Wang}\ \emph
  {et~al.}(2022{\natexlab{c}})\citenamefont {Wang}, \citenamefont {Hu},\ and\
  \citenamefont {Liu}}]{JiaWang2022arXiv2}%
  \BibitemOpen
  \bibfield  {author} {\bibinfo {author} {\bibfnamefont {Jia}\ \bibnamefont
  {Wang}}, \bibinfo {author} {\bibfnamefont {Hui}\ \bibnamefont {Hu}}, \ and\
  \bibinfo {author} {\bibfnamefont {Xia-Ji}\ \bibnamefont {Liu}},\ }\href@noop
  {} {\enquote {\bibinfo {title} {Two-dimensional spectroscopic diagnosis of
  quantum coherence in fermi polarons},}\ } (\bibinfo {year}
  {2022}{\natexlab{c}}),\ \bibinfo {note} {{arXiv}:2207.14509}\BibitemShut
  {NoStop}%
\bibitem [{\citenamefont {Wang}\ \emph
  {et~al.}(2012{\natexlab{a}})\citenamefont {Wang}, \citenamefont {D'Incao},
  \citenamefont {Esry},\ and\ \citenamefont {Greene}}]{Jia2012PRL}%
  \BibitemOpen
  \bibfield  {author} {\bibinfo {author} {\bibfnamefont {Jia}\ \bibnamefont
  {Wang}}, \bibinfo {author} {\bibfnamefont {J.~P.}\ \bibnamefont {D'Incao}},
  \bibinfo {author} {\bibfnamefont {B.~D.}\ \bibnamefont {Esry}}, \ and\
  \bibinfo {author} {\bibfnamefont {Chris~H.}\ \bibnamefont {Greene}},\
  }\bibfield  {title} {\enquote {\bibinfo {title} {Origin of the three-body
  parameter universality in efimov physics},}\ }\href@noop {} {\bibfield
  {journal} {\bibinfo  {journal} {Phys. Rev. Lett.}\ }\textbf {\bibinfo
  {volume} {108}},\ \bibinfo {pages} {263001} (\bibinfo {year}
  {2012}{\natexlab{a}})}\BibitemShut {NoStop}%
\bibitem [{\citenamefont {Wang}\ \emph
  {et~al.}(2012{\natexlab{b}})\citenamefont {Wang}, \citenamefont {Wang},
  \citenamefont {D'Incao},\ and\ \citenamefont {Greene}}]{YujunPRL2012}%
  \BibitemOpen
  \bibfield  {author} {\bibinfo {author} {\bibfnamefont {Yujun}\ \bibnamefont
  {Wang}}, \bibinfo {author} {\bibfnamefont {Jia}\ \bibnamefont {Wang}},
  \bibinfo {author} {\bibfnamefont {J.~P.}\ \bibnamefont {D'Incao}}, \ and\
  \bibinfo {author} {\bibfnamefont {Chris~H.}\ \bibnamefont {Greene}},\
  }\bibfield  {title} {\enquote {\bibinfo {title} {Universal three-body
  parameter in heteronuclear atomic systems},}\ }\href@noop {} {\bibfield
  {journal} {\bibinfo  {journal} {Phys. Rev. Lett.}\ }\textbf {\bibinfo
  {volume} {109}},\ \bibinfo {pages} {243201} (\bibinfo {year}
  {2012}{\natexlab{b}})}\BibitemShut {NoStop}%
\bibitem [{\citenamefont {Wang}\ \emph
  {et~al.}(2012{\natexlab{c}})\citenamefont {Wang}, \citenamefont {D'Incao},
  \citenamefont {Wang},\ and\ \citenamefont {Greene}}]{JiaWangPRA2012}%
  \BibitemOpen
  \bibfield  {author} {\bibinfo {author} {\bibfnamefont {Jia}\ \bibnamefont
  {Wang}}, \bibinfo {author} {\bibfnamefont {J.~P.}\ \bibnamefont {D'Incao}},
  \bibinfo {author} {\bibfnamefont {Yujun}\ \bibnamefont {Wang}}, \ and\
  \bibinfo {author} {\bibfnamefont {Chris~H.}\ \bibnamefont {Greene}},\
  }\bibfield  {title} {\enquote {\bibinfo {title} {Universal three-body
  recombination via resonant $d$-wave interactions},}\ }\href@noop {}
  {\bibfield  {journal} {\bibinfo  {journal} {Phys. Rev. A}\ }\textbf {\bibinfo
  {volume} {86}},\ \bibinfo {pages} {062511} (\bibinfo {year}
  {2012}{\natexlab{c}})}\BibitemShut {NoStop}%
\end{thebibliography}%

\end{document}